\documentclass{article}
\usepackage{arxiv}

\usepackage[utf8]{inputenc}
\usepackage[T1]{fontenc}

\usepackage{amsthm,amsmath,amssymb,amsfonts}
\usepackage{bm}
\usepackage{dsfont}

\usepackage{booktabs}
\usepackage{threeparttable}

\usepackage{nicefrac}
\usepackage{microtype}
\usepackage{setspace}
\usepackage{lineno}
\usepackage{graphicx}
\usepackage{xcolor} 
\usepackage{subcaption}
\usepackage{lscape}
\usepackage{rotating}
\usepackage{algorithm}
\usepackage{algpseudocode}
\usepackage{epigraph}

\usepackage[backend=biber,style=apa,natbib=true]{biblatex}

\usepackage{url}
\usepackage{doi} 
\usepackage{hyperref}
\usepackage{orcidlink}

\title{Principal Covariate Regression with Nuclear Norm Penalty}

\author{
  Kaiwen Liu \orcidlink{0009-0002-7778-0706},
  Lisa Verbeij \orcidlink{0000-0002-8999-6857}, 
  Wouter Weeda \orcidlink{0000-0001-8619-2645}, 
  Mark de Rooij \orcidlink{0000-0001-7308-6210}\\
  Unit Methodology and Statistics, Institute of Psychology\\
  Leiden University\\
  Wassenaarseweg 52, 2333 AK Leiden\\
  \texttt{k.liu.10@fsw.leidenuniv.nl} \\
}

\date{}

\begin{document}
\maketitle
\begin{abstract}
In high-dimensional data settings, dimensionality reduction or variable selection are key steps when using statistical learning techniques. Principal Covariate Regression-type methods aim to perform both dimensionality reduction and (regularized) regression steps in one analysis. However, existing PCovR methods cannot simultaneously select dimensionalities and estimate regularized coefficients, forcing researchers to make ad-hoc choices in the order of these steps. In this study, we propose a novel method called Principal Covariate Regression with Nuclear Norm Penalty (PcovRnnp) that allows simultaneous dimension reduction and estimation of regularized coefficients. 
\end{abstract}



\newcommand{\mdelta}{\mathbf{\Delta}}
\newcommand{\mdeltat}{\mathbf{\Delta}'}
\newcommand{\cx}{c_{x}}
\newcommand{\cy}{c_{y}}

\newcommand{\Q}{\mathbf{Q}}
\newcommand{\Qt}{\mathbf{Q}'}
\newcommand{\K}{\mathbf{K}}
\newcommand{\Kt}{\mathbf{K}'}

\newcommand{\E}{\mathbf{E}}
\newcommand{\Et}{\mathbf{E}'}
\newcommand{\X}{\mathbf{X}}
\newcommand{\Xt}{\mathbf{X}'}
\newcommand{\Y}{\mathbf{Y}}
\newcommand{\Yt}{\mathbf{Y}'}
\newcommand{\z}{\mathbf{z}}
\newcommand{\Z}{\mathbf{Z}}
\newcommand{\Zt}{\mathbf{Z}'}
\newcommand{\Zhat}{\hat{\mathbf{Z}}}
\newcommand{\zhat}{\hat{\mathbf{z}}}

\newcommand{\Xinput}{\X_{\text{input}}}
\newcommand{\Yinput}{\Y_{\text{input}}}
\newcommand{\Xnew}{\X_{\text{new}}}
\newcommand{\Yhatnew}{\hat{\Y}_{\text{new}}}

\newcommand{\mx}{\mathbf{m}_{x}}
\newcommand{\my}{\mathbf{m}_{y}}

\newcommand{\xbar}{\bar{\mathbf{x}}}
\newcommand{\ybar}{\bar{\mathbf{y}}}

\newcommand{\Sx}{\mathbf{S}_{x}}
\newcommand{\Sxinv}{\mathbf{S}_{x}^{-1}}
\newcommand{\Sy}{\mathbf{S}_{y}}
\newcommand{\Syinv}{\mathbf{S}_{y}^{-1}}

\newcommand{\XtXsqrt}{(\mathbf{\Xt\X})^{\frac{1}{2}}}
\newcommand{\XtXsqrtinv}{(\mathbf{\Xt\X})^{-\frac{1}{2}}}

\newcommand{\A}{\mathbf{A}}
\newcommand{\B}{\mathbf{B}}
\newcommand{\Bplus}{\mathbf{B}^{+}}
\newcommand{\Bx}{\mathbf{B}_x}
\newcommand{\By}{\mathbf{B}_y}
\newcommand{\Byt}{\mathbf{B}'_y}
\newcommand{\Bols}{\mathbf{B}_{\text{ols}}}
\newcommand{\Bhat}{\hat{\mathbf{B}}}
\newcommand{\Btilde}{\tilde{\mathbf{B}}}
\newcommand{\Btildeplus}{\tilde{\mathbf{B}}^{+}}
\newcommand{\Btildet}{\tilde{\mathbf{B}}'}
\newcommand{\Btildeols}{\tilde{\mathbf{B}}_{\text{ols}}}
\newcommand{\Btildeolst}{\tilde{\mathbf{B}}'_{\text{ols}}}

\newcommand{\w}{\mathbf{w}}
\newcommand{\W}{\mathbf{W}}
\newcommand{\Wt}{\mathbf{W}'}
\newcommand{\T}{\mathbf{T}}
\newcommand{\Tt}{\mathbf{T}'}

\newcommand{\Pz}{\mathbf{P}}
\newcommand{\Pzt}{\mathbf{P}'}
\newcommand{\Px}{\mathbf{P}_x}
\newcommand{\Pxt}{\mathbf{P}'_x}
\newcommand{\Py}{\mathbf{P}_y}
\newcommand{\Pyt}{\mathbf{P}'_y}

\newcommand{\Pygt}{\mathbf{P}'_{y(g)}}
\newcommand{\Ex}{\mathbf{E}_x}
\newcommand{\Ey}{\mathbf{E}_y}
\newcommand{\Eols}{\mathbf{E}_{ols}}
\newcommand{\Enp}{\mathbf{E}_{np}}

\newcommand{\Extest}{\mathbf{E}_{x,\text{test}}}
\newcommand{\Eytest}{\mathbf{E}_{y,\text{test}}}

\newcommand{\U}{\mathbf{U}}
\newcommand{\Utilde}{\tilde{\mathbf{U}}}
\newcommand{\Utildet}{\tilde{\mathbf{U}}'}
\newcommand{\Utildeols}{\tilde{\mathbf{U}}_{\text{ols}}}
\newcommand{\Utildeolst}{\tilde{\mathbf{U}}'_{\text{ols}}}
\newcommand{\Ut}{\mathbf{U}'}
\newcommand{\D}{\mathbf{D}}
\newcommand{\Dtilde}{\tilde{\mathbf{D}}}
\newcommand{\V}{\mathbf{V}}
\newcommand{\Vtilde}{\tilde{\mathbf{V}}}
\newcommand{\Vtildet}{\tilde{\mathbf{V}}'}
\newcommand{\Vtildeols}{\tilde{\mathbf{V}}_{\text{ols}}}
\newcommand{\Vtildeolst}{\tilde{\mathbf{V}}'_{\text{ols}}}
\newcommand{\Vt}{\mathbf{V}'}

\newcommand{\URplus}{\U_{\Rplus}}
\newcommand{\VRplus}{\V_{\Rplus}}
\newcommand{\VtRplus}{\Vt_{\Rplus}}

\newcommand{\bSigma}{\boldsymbol{\Sigma}}
\newcommand{\bSigmaplus}{\boldsymbol{\Sigma}^{+}}
\newcommand{\bSigmatilde}{\tilde{\boldsymbol{\Sigma}}}
\newcommand{\bSigmatildeplus}{\tilde{\boldsymbol{\Sigma}}^{+}}
\newcommand{\bSigmatildeols}{\tilde{\boldsymbol{\Sigma}}_{\text{ols}}}
\newcommand{\Rplus}{{R^{+}}}
\newcommand{\I}{\mathbf{I}}
\newcommand{\IR}{\mathbf{I}_R}
\newcommand{\IS}{\mathbf{I}_S}
\newcommand{\oneN}{\mathbf{1}_{N}}
\newcommand{\oneNstar}{\mathbf{1}_{N_{*}}}
\newcommand{\Nstar}{N_{*}}

\makeatletter
\newcommand{\dims}[1]{\@ifnextchar\bgroup{\@dims{#1}}{\in \mathbb{R}^{#1}}}
\newcommand{\@dims}[2]{\in \mathbb{R}^{#1 \times #2}}
\makeatother

\newcommand{\lmin}{\lambda_{min}}
\newcommand{\lse}{\lambda_{1se}}
\newcommand{\mumin}{\mu_{\min}}

\newcommand{\tr}{\text{tr}}

\newcommand{\all}{\,:}
\newcommand{\submat}[3]{#1\!\left[#2 ,\, #3 \right]}

\section{Introduction}

Researchers across the social, behavioral, and biological sciences often collect data with many variables measured on the same individuals and face the dual challenge of summarizing key information from the data while still predicting meaningful outcomes. This challenge arises even with more modest datasets, such as when predicting psychological symptoms from questionnaire items, where the goal is to summarize the underlying structure of the questionnaire and use that structure to predict psychological outcomes.
Principal Covariate Regression \citep[PCovR;][]{de_jong_principal_1992} is a statistical learning method that simultaneously reduces the dimensionality of a predictor matrix and predicts an outcome matrix through a set of latent components. Specifically, PCovR balances between principal component analysis \citep[PCA;][]{hotelling_analysis_1933}, and reduced rank regression \citep[RRR;][]{anderson_estimating_1951, izenman_reduced-rank_1975, davies_procedures_1982, tso_reduced-rank_1981, van_den_wollenberg_redundancy_1977}. As a result, the components are both interpretable and predictively informative. The regression coefficients also tend to be more stable than those from ordinary least squares \citep{de_jong_principal_1992}.

Despite these strengths, existing PCovR methods do not simultaneously select the number of components and estimated regularized coefficients. Unpenalized PCovR \citep{de_jong_principal_1992, vervloet_pcovr_2015, gvaladze_pcovr2_2021} typically selects the number of components through cross-validation, but the model coefficients do not have any further regularization and are therefore prone to overfitting, particularly in high-dimensional settings where predictors outnumber observations \citep{van_deun_obtaining_2018}. Penalized PCovR applies ridge or lasso penalties to address overfitting \citep{park_extending_2023, van_deun_obtaining_2018}, but they require the number of components to be specified beforehand. In practice, the true component structure, such as the factor structure in psychological questionnaires or fMRI data, is rarely known, making these methods suboptimal in practice. Furthermore, common component selection heuristics are often insufficient to specify the optimal number of components. Scree plots introduce researcher subjectivity, and exhaustive grid searches for the number of components are computationally costly.

To address the abovementioned limitations, we introduce Principal Covariate Regression with Nuclear Norm Penalty (PcovRnnp). The nuclear norm of a matrix is the sum of its singular values and acts as a convex approximation to the number of components \citep{negahban_estimation_2011, bunea_optimal_2011}. PcovRnnp achieves component selection by using soft-thresholding to discard weak components whose signals fall below the penalty threshold \citep{lu_convex_2012, yuan_dimension_2007}. Furthermore, PcovRnnp regularizes the coefficients by penalizing the magnitude of retained strong components \citep{koltchinskii_nuclear-norm_2011, zhou_regularized_2014}. 

The paper is organized as follows. In Section~\ref{sec:methods}, we present the theoretical derivation of the PcovRnnp method and introduce an efficient algorithm to find the optimal solution. 
In Section~\ref{sec:sim1} and Section~\ref{sec:sim2}, we conduct simulation studies to evaluate the performance of PcovRnnp. 
We then apply PcovRnnp to show its usage in practice: In Section~\ref{sec:app1} and Section~\ref {sec:app2}, we illustrate how PcovRnnp predicts depression levels from coping scores and how PcovRnnp predicts cognitive performance from brain imaging data.  
Finally, we conclude the paper with a brief discussion in Section~\ref{sec:discussion}.

\section{Principal Covariate Regression with Nuclear Norm Penalty}
\label{sec:methods}
\subsection{Notations}
We will use the following notations: matrices are denoted by bold uppercases, matrix transpose by the superscript~$'$, matrix trace by the text tr, vectors by bold lowercase, and scalars by lowercase italics. The indices
$n = 1,\ldots, N$ represents the number of observations, $p = 1,\ldots, P$ represents the number of predictors, $q = 1, \ldots, Q$ represents the number of outcomes, $R$ represents the rank of the component matrix or the number of components, and $r = 1, \ldots, R$ represents the index for each component. We will use the terms penalty and regularization interchangeably. 

Let $\X\dims{N}{P}$ denote the predictor matrix with $N$ observations and $P$ predictors, and $\Y\dims{N}{Q}$ represent the response matrix with $N$ observations and $Q$ response variables. The vector $\ybar 
= [\bar{y}_{1}, \ldots,\bar{y}_{Q}]'
\dims{Q}$ contains column means of $\Y$, and the vector $\xbar  
= [\bar{x}_{1}, \ldots,\bar{x}_{P}]'
\dims{P}$ contains colmn means of $\X$. The diagonal matrix $\Sy = \text{diag}(s_{y_1},\ldots, s_{y_Q}) \dims{Q}{Q}$ containins column-wise standard deviations of $\Y$, and the diagonal matrix$\Sx = \text{diag}(s_{x_1},\ldots, s_{x_P})\dims{P}{P}$ contains column-wise standard deviations of $\X$.
For simplicity, we assume all variables are scaled with a mean of 0 and a variance of 1. That is, given input data $\Xinput$ and $\Yinput$, we standardize the data as $\X = (\Xinput - \oneN\xbar')\Sxinv$ and $\Y = (\Yinput - 
\oneN\ybar')\Syinv$.

\subsection{General PCovR Framework}

To illustrate the rationale of PcovRnnp, we start with the general idea of PCovR and extend the method to PcovRnnp.
PCovR predicts outcomes based on a set of principal covariates, which are low-dimensional subspaces of the space spanned by the columns of the predictor matrix $\X$:
\begin{align}
    \X &= \X\W\Pxt + \Ex = \T\Pxt + \Ex; \label{eq:pca_formal}\\
    \Y &= \X\W\Pyt + \Ey = \T\Pyt + \Ey, \label{eq:rrr_formal}
\end{align}

where $\W\dims{P}{R}$ is the weight matrix, $\Px\dims{P}{R}$ is the orthogonal loading matrix on predictors ($\Pxt\Px$ is diagonal), $\Py\dims{Q}{R}$ is the orthogonal loading matrix on outcomes ($\Pyt\Py$ is diagonal), and $\T\dims{N}{R} = \X\W$ is the component matrix with $R$ orthonormal components coming from the linear combinations of columns in $\X$ with corresponding weight $\W$: $\Tt\T = \IR$, with $\IR$ an identity matrix of rank $R$.

PCovR aims to simultaneously minimize reconstruction error on the predictor matrix $\X$ (through the PCA component, see Equation~\ref{eq:pca_formal}) and prediction error on the outcome matrix $\Y$ (through the RRR component, see Equation~\ref{eq:rrr_formal}). In PCovR, the following objective function is minimized with three sets of parameters ($\W$, $\Px$, and $\Py$):
\begin{equation}
    \mathcal{L}(\W, \Px, \Py) 
    = (1-\alpha) \frac{\|\Y - \X\W\Pyt\|_F^2}{\|\Y\|_F^2} 
    + \alpha \frac{\|\X - \X\W\Pxt\|_F^2}{\|\X\|_F^2}, 
    \label{eq:PCovR_loss}
\end{equation}

where $\|\cdot\|_F$ denotes the Frobenius norm and $\alpha \in [0,1]$ is the weight hyperparameter balancing the prediction on outcome $\Y$ and reconstruction on the predictor matrix $\X$. $\alpha = 0$ represents RRR, whereas $\alpha = 1$ represents PCA. Previous research suggests that optimal $\alpha$ can be estimated using a maximum-likelihood procedure \citep{vervloet_selection_2013, wilderjans_simultaneous_2011}.

To simplify further derivation, we define normalized weights $w_1 = \sqrt{1 - \alpha}/\|\Y\|_F$ and $w_2 = \sqrt{\alpha}/\|\X\|_F$ following \citet{heij_forecast_2007, van_deun_obtaining_2018}  to construct a concatenated response matrix $\Z$ and a joint coefficient matrix $\B$ partitioned into predictition coefficients $\By\dims{P}{Q} = \W\Pyt$ and reconstruction coefficients $\Bx\dims{P}{P} = \W\Pxt$:

\begin{align}
    \Z\dims{N}{(Q+P)} &= [w_1 \Y \quad w_2 \X], \\
    \B\dims{P}{(Q+P)} &= [w_1 \By \quad w_2 \Bx] = [\W] [w_1\Pyt \quad w_2\Pxt] = [\W][\mathbf{P}'].
    \label{eq:B=WPt}
\end{align}

\subsection{Nuclear Norm Penalty, Component Selection, and Coefficient Regularization}
We incorporate a nuclear norm penalty into the PCovR framework to form our PcovRnnp method. In standard settings, the nuclear norm of a matrix $\B$, denoted as $\|\B\|_*$, is defined as the sum of its singular values \citep{negahban_estimation_2011, bunea_optimal_2011, koltchinskii_nuclear-norm_2011}. Nuclear norm penalty fundamentally connects dimensionality reduction and regularization: As the penalty parameter $\lambda$ increases, more singular values of the coefficient matrix are driven toward zero. This results in a lower-rank approximation that captures only the most significant latent dimensions \citep{koltchinskii_nuclear-norm_2011}. 

\subsubsection{Theoretical Properties}

\paragraph{Rank Recovery and Consistency}
The nuclear norm penalty estimator adapts to the intrinsic low-rank structure of $\B$ and performs nearly as well as if the true low-rank structure of $\B$ were known in advance \citep{koltchinskii_nuclear-norm_2011, candes_exact_2009}. Given an appropriate regularization parameter $\lambda$, the nuclear norm penalty estimator can consistently recover the true rank of the underlying coefficient matrix $\B$ with probability close to 1 \citep{yuan_dimension_2007, negahban_estimation_2011}.

\paragraph{Prediction Performance}
The nuclear norm penalty estimator inherits prediction consistency guarantees from nuclear norm penalized regression, providing theoretical error bounds that ensure reliable performance in high-dimensional settings \citep{zhou_regularized_2014, lu_convex_2012}.

\paragraph{Convexity and Global Optimum}
The nuclear norm penalty estimator leads to a convex optimization problem that can be solved efficiently using soft-thresholding on the singular values \citep{cai_singular_2010, lu_convex_2012}. The convexity ensures that the global optimum is attainable.

\subsection{Nuclear Norm under Orthonormality Constraint}
While the standard nuclear norm provides a strong theoretical foundation, interpretability in PCovR framework requires the component scores $\T = \X\W$ to be orthonormal, i.e., $\Tt\T = \Wt\Xt\X\W = \IR$. This orthonormality constraint implies that the nuclear norm should not be evaluated using standard SVD, but rather with respect to the metric $\Xt\X$ \citep{takane_constrained_2016}. This requirement motivates the use of the Generalized Singular Value Decomposition \citep[GSVD;][]{abdi_singular_2007}.

Under GSVD framework, we define the constrained nuclear norm as $\|\B\|_{\dagger} = \sum \sigma_r(\B, \Xt\X)$, which is equivalent to the standard nuclear norm of the auxiliary matrix $\Btilde = (\Xt\X)^{\frac{1}{2}}\B$ \citep{abdi_singular_2007}. We also introduce a normalizing constant $\frac{1}{2}$ to simplify the computation in later steps. The objective function for PcovRnnp could then be expressed as:
\begin{equation}
    \mathcal{L}(\B) = \frac{1}{2}\|\Z - \X\B \|_F^2 + \lambda \|\B\|_{\dagger}.
    \label{eq:constrained_loss}
\end{equation}
where $\lambda \geq 0$ is the regularization parameter that controls the trade-off between model fit and component selection. In this constrained formulation, the rank of the solution $\hat{\B}_\lambda$ identifies the optimal number of selected components, and the value of $\hat{\B}_\lambda$ corresponds to the regularized coefficients.

\subsubsection{Solving the Optimization Problem}

To solve Equation~\ref{eq:constrained_loss}, we introduce the ordinary least squares (OLS) estimator $\Bols = (\Xt\X)^{-1}\Xt\Z$ and its auxiliary matrix $\Btildeols = (\Xt\X)^{\frac{1}{2}}\Bols$. By decomposing the squared error term, using OLS properties \citep{seber_linear_2003}, and recalling the equivalence between the $\|\B\|_{\dagger}$ and $\|\Btilde\|_{*}$, the loss function simplifies into:
\begin{equation}
    \mathcal{L}(\Btilde) = \frac{1}{2}\|\Btildeols - \Btilde\|_F^2 + \lambda \|\Btilde\|_{*} + \text{const},
    \label{eq:simplified_loss}
\end{equation}
which is an unconstrained nuclear norm minimization problem. According to von Neumann's trace inequality \citep{carlsson_von_2021}, the optimal $\Btilde$ must share its left and right singular vectors with $\Btildeols$. Let $\Btildeols = \Utildeols\bSigmatildeols\Vtildeolst$ be the SVD of $\Btildeols$ with $r = 1 \ldots R$ singular values $\{\tilde{\sigma}_{r(\text{ols})}\}$. The optimal solution can be obtained via soft-thresholding over the singular values \citep{cai_singular_2010}:
\begin{equation}
    \tilde{\sigma}_r^+ = \max(0, \tilde{\sigma}_{r(\text{ols})} - \lambda).
    \label{eq:soft_thresholding}
\end{equation}

Let $R^+ \leq R$ be the number of non-zero singular values after soft-thresholding. We construct the regularized diagonal matrix $\bSigmaplus \in \mathbb{R}^{R^+ \times R^+}$ containing these $R^+$ non-zero singular values. The final optimal coefficient matrix $\Bplus$ is constructed as:
\begin{align}
    \URplus &= \text{first } R^+ \text{ columns of } \XtXsqrtinv \Utildeols; \nonumber \\
    \VRplus &= \text{first } R^+ \text{ columns of } \Vtildeols; \nonumber \\
    \Bplus  &= \URplus \bSigmaplus \VtRplus.
\end{align}

This formulation ensures that the selected components simultaneously optimize the prediction objective while maintaining the orthonormality constraints inherent in PCovR, with the number of components automatically determined by the regularization strength $\lambda$.

\subsubsection{Connection to PCovR Parameters}

The PcovRnnp solution naturally connects to the classical PCovR parameterization in terms of the weight matrix $\W$, predictor loadings $\Px$, and response loadings $\Py$ based on Equation~\ref{eq:B=WPt}. The coefficient matrix $\mathbf{B^+}$ has rank $\Rplus$, which represents the number of selected latent components. With this effective dimensionality established, we obtain the PCovR parameterization from the truncated GSVD of $\mathbf{B^+} = \URplus\bSigma^+\VtRplus$:

\begin{align}
    \W &= \URplus;\\
    \T &= \X\URplus, \quad\text{with } \Tt\T = \I_\Rplus;\\
    \Py &= \frac{1}{w_1}\submat{\V_{\Rplus}}{1,\ldots,Q}{\all} \bSigmaplus,\\
        &\qquad \text{where rows } 1:Q \text{ correspond to the } Q \text{ response variables};\nonumber\\
    \Px &= \frac{1}{w_2}\submat{\V_{\Rplus}}{(Q+1):(Q+P)}{\all} \bSigmaplus,\\
        &\qquad \text{where rows } (Q+1):(Q+P) \text{ correspond to the } P \text{ predictor variables.}\nonumber\\
\end{align}


The predictor and response loadings are extracted as specific row blocks from $\V_{\Rplus}$, rescaled, and then multiplied by $\bSigma^+$: $\Py$ uses the first $Q$ rows (associated with the $Q$ response variables) and is rescaled by $\frac{1}{w_1}$, whereas $\Px$ uses the next $P$ rows (associated with the $P$ predictor variables) and is rescaled by $\frac{1}{w_2}$. The construction of $\B$ from $\W$, $\Px$, and $\Py$ can be validated using Equation~\ref{eq:B=WPt}.

\subsubsection{Prediction, Unstandardized Coefficients, and Intercept}

In previous sections, we focused on finding optimized coefficients on standardized data (centered by column mean and scaled by column standard deviation) to simplify many derivations. When data are standardized, the intercept term is always zero since both predictors and responses have a mean of zero. However, in practice it is also important to predict responses in the original scale, and the intercept is often non-zero and needs to be estimated. We first show that predictions can be made on new data, then derive the unstandardized coefficients and intercept.

Denote the new predictor matrix as $\Xnew \dims{\Nstar}{P}$ and the predicted response matrix as $\Yhatnew \dims{\Nstar}{Q}$. 
Suppose the model is fitted on standardized training data and we obtain coefficients $\By$ that predict standardized $\Y$ scores based on standardized $\X$ scores.
For new data, if we standardize them using the training data statistics, the prediction equation is:
$\Yhatnew^{(s)} = \Xnew^{(s)}\By,$
where $\Yhatnew^{(s)} = (\Yhatnew - \oneNstar\ybar')\Syinv$ and $\Xnew^{(s)} = (\Xnew - \oneNstar\xbar')\Sxinv$ are the standardized predictions and predictors, respectively. Here $\xbar$ and $\ybar$ are the column means from the training data, and $\Sx$ and $\Sy$ are diagonal matrices of column standard deviations from the training data.
Substituting the standardization formulas into the prediction equation, we could rearrange the terms into a common prediction form $\Yhatnew = \oneNstar\my' + \Xnew\By$:
\begin{align*}
    \Yhatnew &= \oneNstar(\ybar' - \xbar'\Sxinv\By\Sy) + \Xnew\Sxinv\By\Sy,
\end{align*}
where the intercept vector (length $Q$) is expressed as $\my = \ybar - \Sy(\By)'\Sxinv\xbar$, and the unstandardized coefficient matrix is expressed as $\Sxinv\By\Sy$.

\subsubsection{Tuning Hyperparameter with Cross-Validation}
To find the optimal number of components and corresponding coefficients, we need to tune the hyperparameter $\lambda$. A common approach is to use $K$-fold cross-validation (CV). The dataset $\X$ is partitioned into $K$ approximately equally sized folds. For each fold $k = 1, \ldots, K$, the training data is denoted as $\X^{(\setminus k)}$, which includes all folds except the $k$-th fold. The validation data is denoted as $\X^{(k)}$, which is the held-out $k$-th fold.

For each candidate value of $\lambda$, the model is trained on $\X^{(\setminus k)}$ and predictions are made on $\X^{(k)}$. The validation error is computed using a loss function such as root mean squared error (RMSE). \footnote{We also explored weighted loss on $\Y$ and $\X$ in simulation. Only using loss on $\Y$ yields better perfomance.} The average validation error across all $K$ folds is then calculated:
\begin{equation}
    \text{CV}(\lambda) = \frac{1}{K} \sum_{k=1}^{K} \text{Loss}(\Y^{(k)}, \hat{\Y}^{(k)}_{\lambda}).
\end{equation}

Two commonly used selection criteria are:
\begin{itemize}
    \item $\lambda_{\min}$: the value of $\lambda$ that minimizes the cross-validation error.
    \item $\lambda_{\text{1se}}$: the largest value of $\lambda$ whose error is within one standard error of the minimum error, providing a more parsimonious model with a lower number of components selected.
\end{itemize}

\subsection{Algorithm}
\label{subsec:algorithm}

A summary of our algorithm can be found in Algorithm~\ref{alg:PCovR_nnp} and Algorithm~\ref{alg:PCovR_nnp_cv}. Algorithm~\ref{alg:PCovR_nnp} describes how to obtain the optimal coefficient matrix given a nuclear norm penalty strength $\lambda$, while Algorithm~\ref{alg:PCovR_nnp_cv} extends this to hyperparameter tuning via $K$-fold cross-validation.

\subsubsection{Finding Optimal Coefficients}

Algorithm~\ref{alg:PCovR_nnp} computes the optimal solution given penalty strength $\lambda$. Our approach obtains a closed-form solution based on soft-thresholding on the GSVD of the OLS coefficients. As a result, our approach avoids iterative optimization and ensures computational efficiency.

\begin{algorithm}[htbp]
\caption{PcovRnnp}
\label{alg:PCovR_nnp}
\begin{algorithmic}[1]
\Require Predictor matrix $\X\dims{N}{P}$, response matrix $\Y\dims{N}{Q}$, regularization parameter $\lambda \geq 0$, weight parameter $\alpha$.
\Ensure Coefficient matrix $\B$, Weight matrix $\W$, component scores $\T$, predictor loadings $\Px$, response loadings $\Py$.
\State Compute standardized weight: $w_1$, $w_2$ and build joint response matrix $\Z$.
\State Compute OLS coefficient matrix of $\Z$ on $\X$: $\Bols$.
\State Perform GSVD on $\Bols = \U\bSigma\Vt$ in the metric of $\Xt\X$, $\I$.
\State Apply soft-thresholding on $\bSigma$ with penalty $\lambda$.
\State Update coefficient matrix $\B^+$.
\State Obtain $\W$, $\T$, $\Px$, and $\Py$ from $\B^+$.
\State \Return $\B^+$, $\W$, $\T$, $\Px$, and $\Py$.
\end{algorithmic}
\end{algorithm}

\subsubsection{Finding Optimal Hyperparameter}

Algorithm~\ref{alg:PCovR_nnp_cv} extends Algorithm~\ref{alg:PCovR_nnp} to a cross-validation scheme for tuning the penalty $\lambda$. Our procedure offers 
two key advantages. 
First, the closed-form solution enables efficient computation across the entire regularization path: within each fold, the GSVD is computed only once and then reused to construct solutions for all $\lambda$ values. As a result, for $K$-fold cross-validation, the GSVD only needs to be performed $K$ times, after which soft-thresholding yields solutions for every $\lambda$ on the path. Second, we exploit the property of the nuclear norm penalty by computing a truncated GSVD via the \texttt{RSpectra} package \citep{qiu_rspectra_2016}. Since the nuclear norm penalty shrinks small singular values to zero, only a small number of leading singular values are needed to construct the solution. 
These advantages make PcovRnnp computationally feasible for moderately large datasets without requiring specialized hardware or parallelization to perform cross-validation.
\begin{algorithm}[htbp]
\caption{PcovRnnp Hyperparameter Tuning with CV}
\label{alg:PCovR_nnp_cv}
\begin{algorithmic}[1]
\Require Predictor matrix $\X\dims{N}{P}$, response matrix $\Y\dims{N}{Q}$, weight parameter $\alpha$, number of folds $K$, sequence of regularization parameters $\{\lambda_1, \lambda_2, \ldots, \lambda_M\}$.
\Ensure Optimal regularization parameter $\lambda^*$.
\State Partition $\X$ and $\Y$ into $K$ folds: $\{\X^{(k)}, \Y^{(k)}\}_{k=1}^{K}$.
\For{each fold $k = 1, \ldots, K$}
    \For{each $\lambda_m$ in $\{\lambda_1, \lambda_2, \ldots, \lambda_M\}$}
        \State Construct training data: $\X^{(\setminus k)}$, $\Y^{(\setminus k)}$.
        \State Train PcovRnnp on $\X^{(\setminus k)}$, $\Y^{(\setminus k)}$ with $\lambda_m$ and $\alpha$ using Algorithm~\ref{alg:PCovR_nnp}.
        \State Obtain coefficient matrix $\B_{y,\lambda_m}^{(\setminus k)}$.
        \State Compute predictions on validation fold: $\hat{\Y}^{(k)}_{\lambda_m} = \X^{(k)}\B^{(\setminus k)}_{y,\lambda_m}$.
        \State Compute validation error: $\text{Loss}_k(\lambda_m) = \text{Loss}(\Y^{(k)}, \hat{\Y}^{(k)}_{\lambda_m})$.
    \EndFor
    \State Compute average cross-validation error: $\text{CV}(\lambda_m) = \frac{1}{K} \sum_{k=1}^{K} \text{Loss}_k(\lambda_m)$.
    \State Compute standard error: $\text{SE}(\lambda_m) = \sqrt{\frac{1}{K}\sum_{k=1}^{K}(\text{Loss}_k(\lambda_m) - \text{CV}(\lambda_m))^2}$.
\EndFor
\State Find $\lambda_{\min} = \operatorname*{arg\,min}_{\lambda_m} \text{CV}(\lambda_m)$.
\State Find $\lambda_{\text{1se}} = \max\{\lambda_m : \text{CV}(\lambda_m) \leq \text{CV}(\lambda_{\min}) + \text{SE}(\lambda_{\min})\}$.
\State \Return  optimal $\lambda_{\text{opt}} \in \{\lambda_{\min}, \lambda_{\text{1se}}\}$ based on desired model complexity. The final model can then be trained on the full dataset $\X$, $\Y$ with $\lambda_{\text{opt}}$ and $\alpha$ using Algorithm~\ref{alg:PCovR_nnp}.
\end{algorithmic}
\end{algorithm}
\section{Simulation: Hyperparameter Selection and Benchmarking under Gaussion Noise}
\label{sec:sim1}

The primary objective of this simulation study is to systematically evaluate hyperparameter
selection strategies for PcovRnnp. 
We focus on two hyperparameters. The first is the penalty
parameter $\lambda$. Specifically, we investigate whether $\lmin$ or $\lse$
yields a superior balance between model complexity and predictive accuracy.
The other hyperparameter is the cross-validation criterion used for model selection.
Previous research has predominantly relied on the prediction loss on the validation $\Y$
alone \citep{van_deun_obtaining_2018, park_extending_2023}. However, given that the PCovR framework explicitly balances the reconstruction of both
$\X$ and the prediction of $\Y$ \citep{de_jong_principal_1992}, it is worth investigating whether a weighted
combination of the prediction error on validation $\Y$ and the reconstruction error on
validation $\X$ could improve component recovery and yield more accurate predictions and coefficient recovery.
We evaluate the model performance in three aspects: The prediction error, measured by root mean squared error (RMSE) on the outcome with the held-out test set of size 1000; the recovery error, measured by RMSE on coefficients; and the number of components selected by the model.

The secondary objective is to compare the relative performance of PcovRnnp against an established method. Given the
complexity introduced by examining multiple variants of PcovRnnp, we keep the comparison
straightforward and include Ridge regression \citep{zou_regularization_2005} from package \texttt{glmnet} \citep{friedman_glmnet_2008} as a reference method. We only compare model performance between PcovRnnp and Ridge in terms of the prediction error on outcome and the recovery error on coefficients, because Ridge does not involve component selection.
We expect that PcovRnnp to achieve lower RMSE on outcome and lower RMSE on coefficients because PcovRnnp might use the low-rank structure of the generated data to make more accurate estimations \citep{de_jong_principal_1992}.

\subsection{Design}
Following the simulation workflows established by \citet{park_variable_2024,guerra-urzola_guide_2021}, we conducted a comprehensive Monte Carlo study to evaluate the performance of PcovRnnp across a diverse range of experimental conditions: sample size, number of predictors, number of responses, latent structure, and noise level. Specifically, this workflow aims to evaluate the model performance in an ideal case. Therefore, the simulation noise is designed to strictly follow a Gaussian distribution, which aligns with the theoretical assumptions of PCovR \citep{vervloet_model_2016, de_jong_principal_1992}.

\subsubsection{Parameter Grid}

The simulation parameters were varied systematically across the following grid:

\begin{itemize}
    \item Training sample size: $N_\text{train} \in \{50, 100\}$;
    \item Number of predictor variables: $P \in \{25, 75, 125\}$;
    \item Number of response variables: $Q \in \{1, 5, 10\}$;
    \item Variance Accounted For (VAF): $\text{VAF} \in \{0.50, 0.80, 0.95\}$ for both $\X$ and $\Y$.
   
\end{itemize}

Furthermore, we fixed the following parameters across all simulation conditions:

\begin{itemize}
    \item True rank of latent component matrix $\T$: $R = 4$;
    \item Test sample size: $N_\text{test} = 1000$;
    \item Monte Carlo repetitions: 100 per parameter combination.
\end{itemize}

In total, this design resulted in $2 \times 3 \times 3 \times 3 \times 100 = 5400$ simulation runs.

\subsubsection{Data Generation Process}

The data generation process reflects a latent structure in which both $\X\dims{N}{P}$ and $\Y\dims{N}{Q}$ are driven by a shared set of underlying components $\T\dims{N}{R}$.
We describe each element of this process in turn.

We first draw the latent component matrix $\T$, which represents the true, unobserved structure common to both $\X$ and $\Y$. To ensure that the components carry a strong signal relative to the noise that will be added later, each column of $\T$ is drawn independently from a normal distribution: 
$\T \dims{N}{R} \sim \mathcal{MVN}(0,\, \bSigma = 50^2 \I).$ The variance of $\T$ follows the design from \citet{guerra-urzola_guide_2021}.

We then draw the weight matrix $\W\dims{P}{R}$ and the loading matrix $\Px\dims{P}{R}$ to link the latent component matrix $\T$ to the observed predictor matrix $\X$ as described in Equation~\ref{eq:pca_formal}.
For identification purposes, we impose two constraints as mentioned in \citet{guerra-urzola_guide_2021}: First, the loading matrix $\Px$ equals the weight matrix
$\W\dims{P}{R}$; Second, the weight matrix is orthonormal $\Wt\W = \IR$. These two constraints guarantee that the latent components are correctly linked to the predictors
$\T = \X\W = \T\Wt\W = \T$. In this context, $\W$ is obtained via the QR decomposition of a random matrix with entries drawn from $\mathcal{N}(0,1)$.

Similarly, we draw the loading matrix $\Py\dims{Q}{R}$ to link the latent scores $\T$ to link the latent scores to the observed response matrix $\Y$ as described in Equation~\ref{eq:rrr_formal}. Due to the fact that $Q$, i.e., $\text{max}[\text{rank}(\Pyt\Py)]$, could be smaller than $R$,  no orthonormality constraint is placed on $\Py$. Instead, its entries are drawn independently from a uniform distribution:
$\Py\dims{Q}{R} \leftarrow \mathcal{U}(-1,1).$

Finally, we construct the observed predictor matrix $\X$ and response matrix $\Y$ by combining the latent signal with additive noise. The noise matrices $\Ex\dims{N}{P}$ and $\Ey\dims{N}{Q}$ are independently drawn from standard multivariate normal distributions,
$\Ex\dims{N}{P} \sim \mathcal{MVN}(0,\, \bSigma_{\Ex} = \I), \\
\Ey\dims{N}{Q} \sim \mathcal{MVN}(0,\, \bSigma_{\Ey} = \I),$
and are rescaled by constants $\cx$ and $\cy$ respectively, so that the noise level corresponds to a target VAF criterion. The $\X$ and $\Y$ matrices are then formed as:
\begin{align*}
\X &\leftarrow \T\Wt + \sqrt{\cx}\,\Ex, \\
\Y &\leftarrow \T\Pyt + \sqrt{\cy}\,\Ey.
\end{align*}

Finally, to evaluate out-of-sample predictive performance, test data were generated using the same underlying structure. Specifically, new latent scores $\T_\text{test}\dims{N_\text{test}}{R}$ and noise terms $\Extest$ and $\Eytest$ were drawn independently, while the same loading matrices $\W$ and $\Py$ from the training data generation were reused. This ensures that the test data share the same structural parameters as the training data, while having independent latent components. Results will be aggregated across the 100 Monte Carlo replications for each parameter combination.



\subsection{Results}

In this section, we first examined PcovRnnp's performance (error in predicting outcomes, error in recovering coefficients, and component selection) under different hyperparameter
selection strategies in a setting where both $\X$ and $\Y$ are generated through low-rank latent components plus random Gaussian noise. Two factors were varied within PcovRnnp: (1) the $\lambda$ selection rule ($\lse$ vs.\ $\lmin$), and (2) the cross-validation loss objective (prediction error on $\Y$ only vs.$\Y\X$ a weighted combination of prediction error on $\Y$ and reconstruction error on $\X$). We refer to these as the $\Y$-loss and $\Y\X$-loss variants, respectively.
We then examined the performance of PcovRnnp against Ridge regression in the same setting.

All figures in this section follow a consistent layout: The columns are grouped by the number of responses $(Q = 1, 5, 10)$, forming three blocks. The first block refers to univariate responses whereas the other two block refer to multivariate responses. Inside each block, the left panel displays results aggregated from conditions where the number of observations is more than the number of predictors ($N > P$). In contrast, the right panel displays results aggregated from conditions where the 
number of observations is less than the number of predictors ($N < P$). The rows refer to VAF levels. The legend denotes different methods and their hyperparameter selection rules in the format Method(lambda choice, loss objective (for pcovrnnp only).
The legend identifies each method along with its hyperparameter selection rule, formatted as Method($\lambda$ choice, loss objective), where the loss objective applies only to PcovRnnp.

\newpage

\begin{figure}[htbp]
  \centering
  \includegraphics[width=1\linewidth]{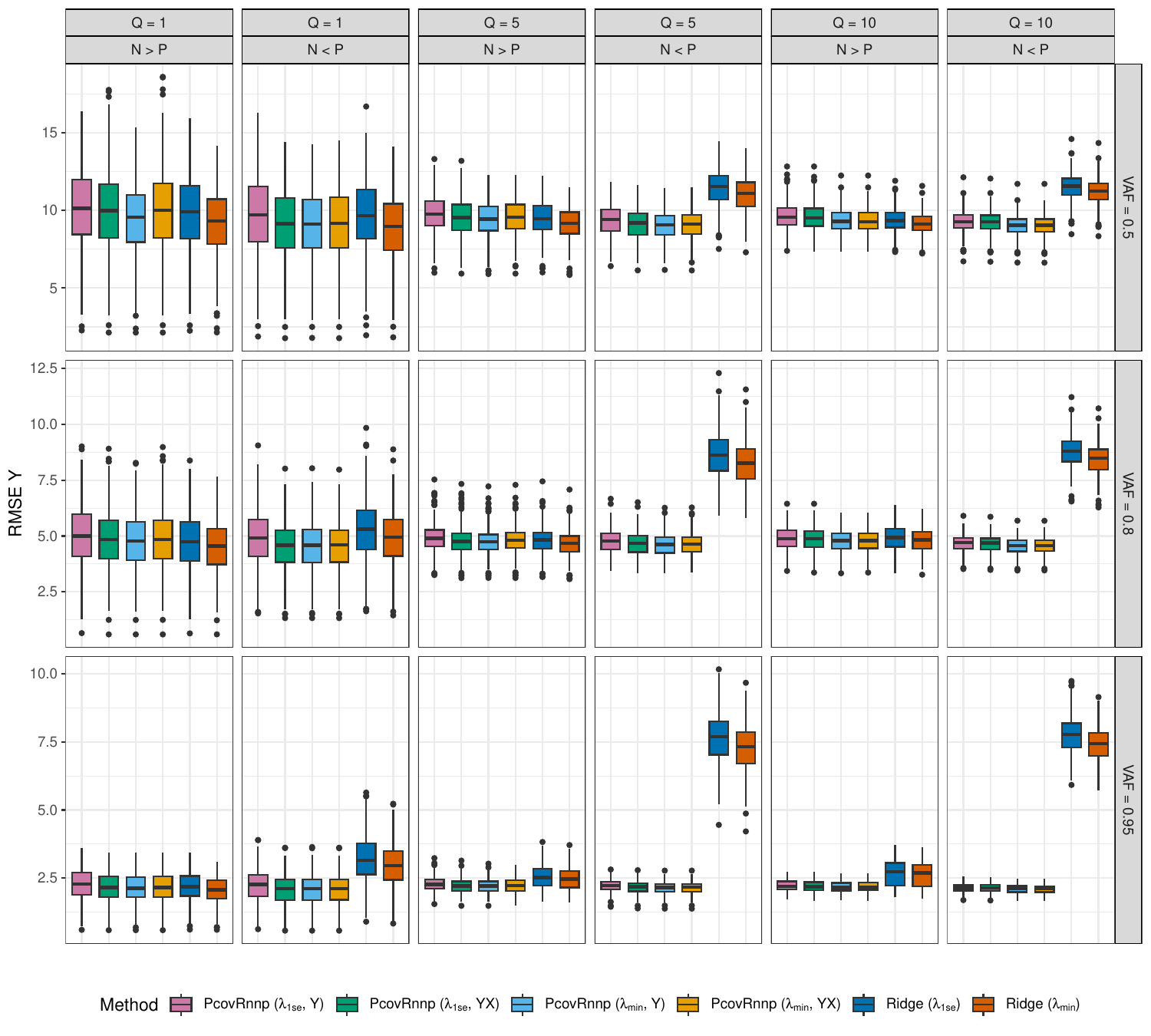}
  \caption{\textit{Root Mean Squared Error of $Y$ in Simulation~1.}}
  \label{fig:sim1_rmse_y}
  \vspace{0.2cm}
  \begin{minipage}{\linewidth}
    \small
    \textit{Note.} The legend denotes different methods and their hyperparameter
    selection rules. PcovRnnp($\lse$,Y/YX): PcovRnnp using the 1-standard-error rule ($\lse$) targeting only $\Y$ or both $\Y$ and $\X$
    as CV objectives. PcovRnnp($\lmin$,Y/YX): PcovRnnp using the minimum
    CV error rule ($\lmin$) targeting only $\Y$ or both $\Y$ and $\X$.
    Ridge($\lse$)/Ridge($\lmin$): Ridge regression using $\lambda_{1\text{se}}$ and
    $\lambda_{\text{min}}$ rules. 
    Test sample size $N_{\text{test}} = 1000$.
    Results are averaged over $100$ repetitions.
  \end{minipage}
\end{figure}

Figure~\ref{fig:sim1_rmse_y} shows the results regarding model performance in predicting outcomes. Regarding prediction error, nearly all PcovRnnp variants achieve lower RMSE on $\Y$ than Ridge regression in high-dimensional settings ($N < P$) with multivariate responses. This predictive advantage becomes stronger as VAF increases. Conversely, in low-dimensional settings ($N > P$) or high-dimensional settings with a univariate response, PcovRnnp shows comparable predictive accuracy to Ridge, with negligible differences between the $\lse$ and $\lmin$ variants.

\newpage

\begin{figure}[htbp]
  \centering
  \includegraphics[width=1\linewidth]{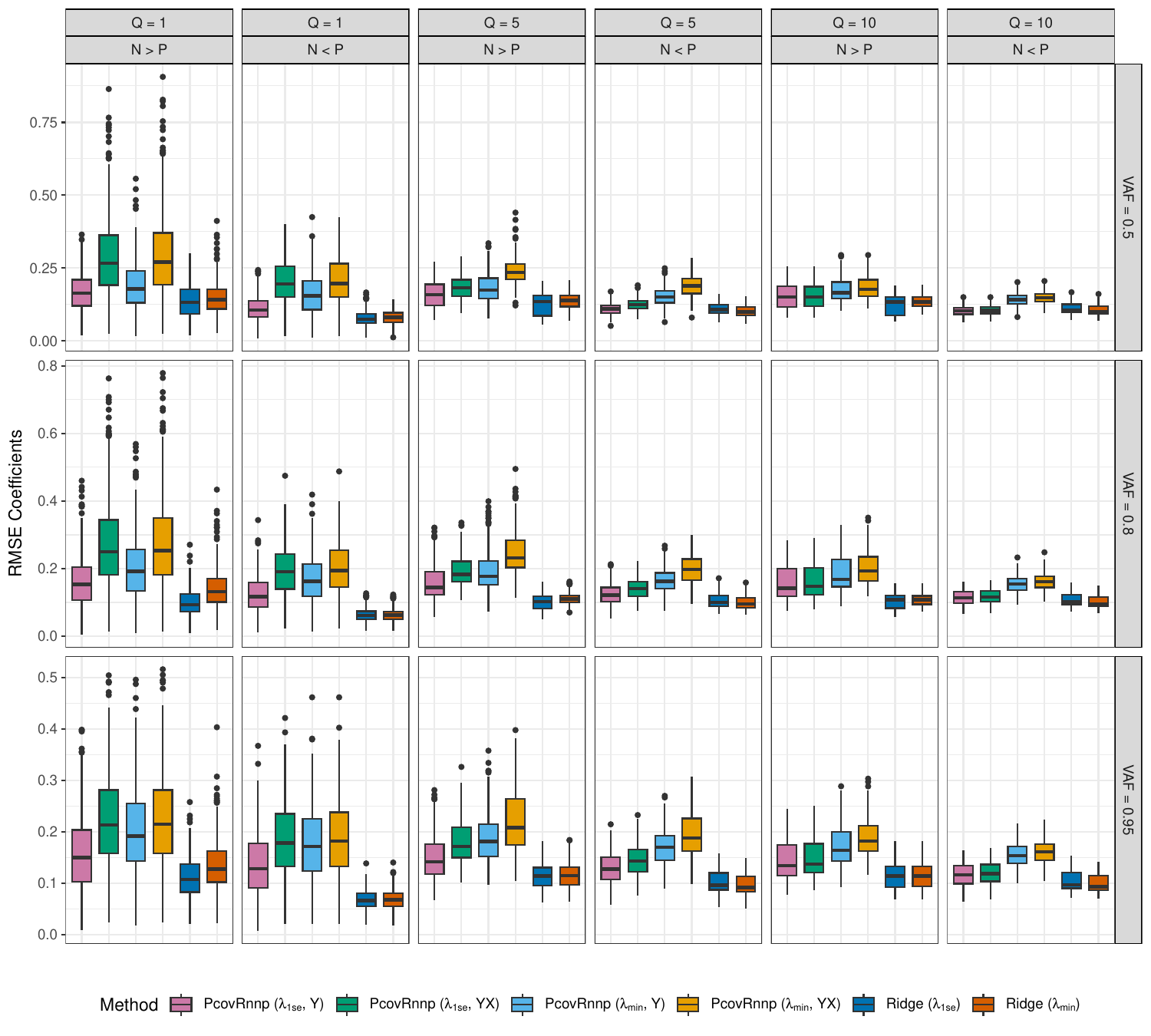}
  \caption{\textit{Root Mean Squared Error of Coefficients in Simulation~1.}}
  \label{fig:sim1_rmse_coef}
  \vspace{0.2cm}
  \begin{minipage}{\linewidth}
    \small
    \textit{Note.} The legend denotes different methods and their hyperparameter
    selection rules. PcovRnnp($\lse$,Y/YX): PcovRnnp using the 1-standard-error rule ($\lse$) targeting only $\Y$ or both $\Y$ and $\X$
    as CV objectives. PcovRnnp($\lmin$,Y/YX): PcovRnnp using the minimum
    CV error rule ($\lmin$) targeting only $\Y$ or both $\Y$ and $\X$.
    Ridge($\lse$)/Ridge($\lmin$): Ridge regression using $\lambda_{1\text{se}}$ and
    $\lambda_{\text{min}}$ rules. 
    Test sample size $N_{\text{test}} = 1000$.
    Results are averaged over $100$ repetitions.
  \end{minipage}
\end{figure}

Figure~\ref{fig:sim1_rmse_coef} shows the results regarding model performance in recovering coefficients. PcovRnnp ($\lse$) consistently achieves lower RMSE on the regression coefficients than PcovRnnp ($\lmin$), particularly in scenarios with multivariate responses. This suggests that the more conservative $\lambda$ selection of $\lse$ yields a sparser and more accurate coefficient structure, even when predictive performance on $\Y$ is similar across variants.

\newpage

\begin{figure}[htbp]
  \centering
  \includegraphics[width=1\linewidth]{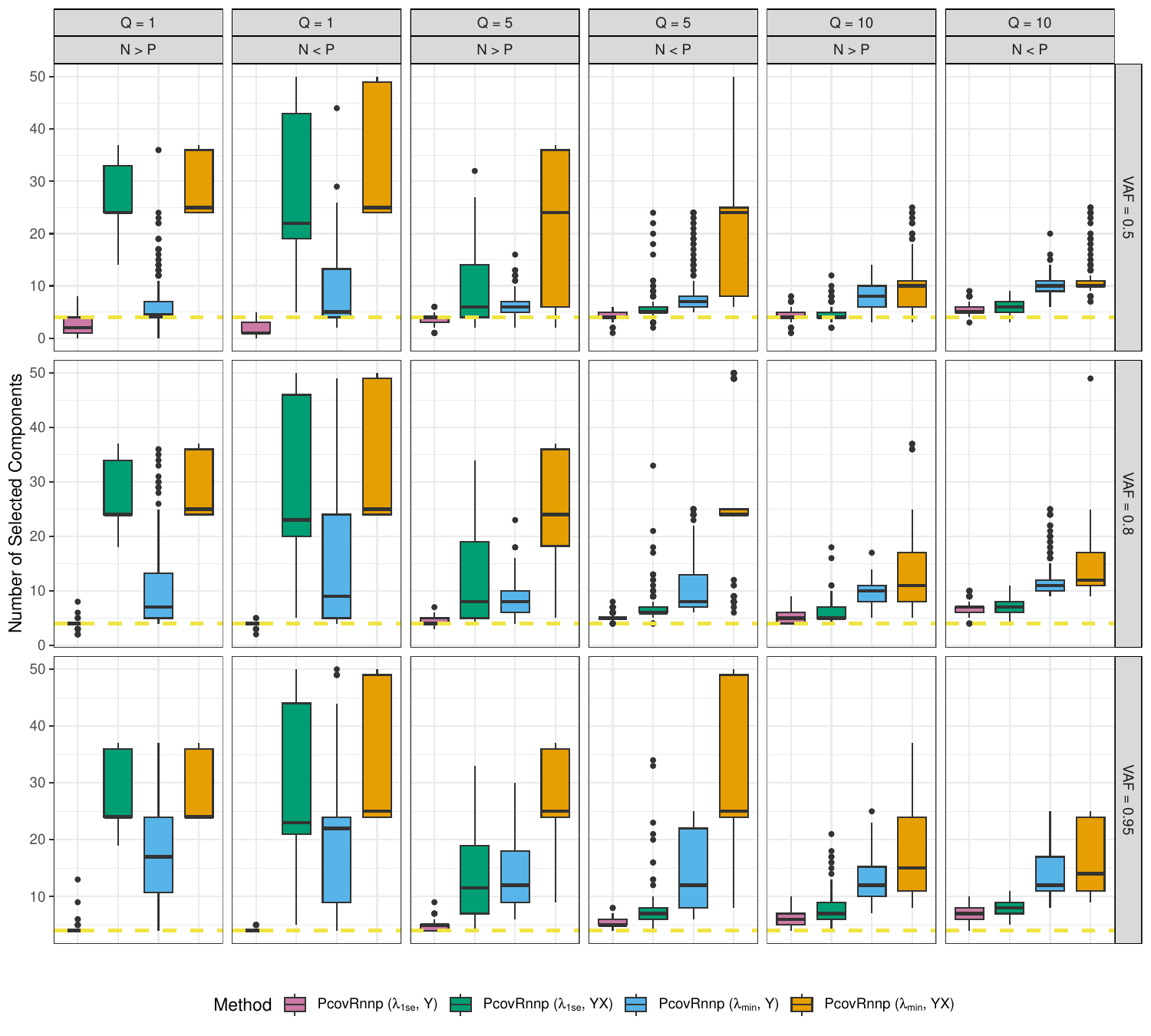}
  \caption{\textit{Number of Components Selected by PcovRnnp in Simulation~1.} }
  \label{fig:sim1_ncomp}
  \vspace{0.2cm}
  \begin{minipage}{\linewidth}
    \small
    \textit{Note.} The legend denotes different methods and their hyperparameter
    selection rules. PcovRnnp($\lse$,Y/YX): PcovRnnp using the 1-standard-error rule ($\lse$) targeting only $\Y$ or both $\Y$ and $\X$
    as CV objectives. PcovRnnp($\lmin$,Y/YX): PcovRnnp using the minimum
    CV error rule ($\lmin$) targeting only $\Y$ or both $\Y$ and $\X$.
    Test sample size $N_{\text{test}} = 1000$.
    The horizontal dashed line refers to True rank $R = 4$.
    Results are averaged over $100$ repetitions.
  \end{minipage}
\end{figure}

Figure~\ref{fig:sim1_ncomp} shows the results regarding component selection among PcovRnnp variants. All PcovRnnp variants tend to overselect the number of components relative to the true rank. However, the degree of overselection differs across variants: $\lse$ selects more accurately to true rank than $\lmin$, and the $\Y$-loss objective selects more accurately to true rank than the $\Y\X$-loss objective. The combination of $\lse$ and the $\Y$-loss criterion yields the most accurate component selection overall.

\section{Simulation: PcovRnnp's Performance on Rank-Noisy Predictor Matrix}
\label{sec:sim2}
In this second simulation, we examine whether PcovRnnp can correctly identify the true rank of the predictive signal when $\X$ contains non-predictive latent structure beyond the signal subspace.
In practice, predictor matrices rarely follow the idealized low-rank structure with purely Gaussian noise as assumed in Section~\ref{sec:sim1}. As noted by \citet{zajac_eigenvalue_2013}, empirical $\X$ matrices (such as obtained from fMRI data) typically exhibit a pronounced singular value structure: a few dominant components capture the true signal, while a long tail of smaller singular values reflects measurement noise or weak latent factors. Because noise is embedded within the singular spectrum rather than separated from it, rank selection becomes more challenging. Thus, a PcovRnnp model might tend to select non-predictive components. 

Based on the results from Section~\ref{sec:sim1}, we set PcovRnnp to use the 1-standard-error rule ($\lse$) targeting only $\Y$ as CV objective. We also set Ridge to use the 1-standard-error rule ($\lse$) because Ridge with ($\lse$) does not significantly underperform Ridge with $\lmin$ and the $\lse$ approach is recommended in common practice \citep{friedman_glmnet_2008}.
Regarding predictive performance, we expect PcovRnnp to achieve lower RMSE on outcomes than Ridge in high-dimensional settings. We expect PcovRnnp to achieve similar RMSE on regression coefficients compared to Ridge in high-dimensional settings. Regarding rank selection, we expect PcovRnnp to recover a rank close to the true rank $R = 4$ when the variance accounted for (VAF) is high, but to overselect ranks when VAF is low, as the predictive components become harder to distinguish from non-predictive components.

\subsection{Design}

\subsubsection{Parameter Grid}

We used the same parameters as in Section~\ref{sec:sim1} except VAF: We adjust the VAF level to 0.5, 0.7, and 0.9 to ensure that noises are detectable given two types of noise (structured and random) in this simulation. We added a new fixed parameter: the number of non-predictive components $S = 6$. This allows us to experiment with the model performance when the noise is related to the latent structure of $\X$. 

\subsubsection{Data Generation Process}

We used the same data generation pattern for $\Px$, $\Py$, and $\Y$. The key difference from Section~\ref{sec:sim1} is that the predictor matrix $\X$ now contains both a signal component $\T\dims{N}{R}$ and a non-predictive component $\Q\dims{N}{S}$ in $\X$'s latent structure. 
For simplicity, the signal and non-predictive components of $\X$ are restricted to be orthogonornal. 
We first present that
$\X$ is constructed as $\X \leftarrow \T\Wt + \sqrt{c}\Q\Kt + \sqrt{\cx}\Ex$, where $c$ and $\cx$ are scaling constants to achieve the target VAF. Then, we explain how to construct $\X$ using the following procedures.

To ensure the simulation is controllable and identifiable, we define how VAF for $\X$ is calculated:
\begin{equation}\label{eq:vaf_x}
  \mathrm{VAF}_X
  \;=\; \frac{\|\T\Wt\|_F^2}
               {\|\T\Wt\|_F^2
                + c\|\Q\Kt\|_F^2
                + \cx\|\Ex\|_F^2}.
\end{equation}
Furthermore, we set the relationship $cx = 0.25c$, ensuring that $80\%$ of the noise in $\X$ comes from $\Q$, reflecting the situation that structured noise dominates random measurement error. By defining $\delta = (1 - \mathrm{VAF}_X)/\mathrm{VAF}_X$, the target $\mathrm{VAF}_X$ is achieved by setting:
$c\|\Q\|_F^2 = 0.8\delta\|\T\|_F^2, \quad \cx\|\Ex\|_F^2 = 0.2\delta\|\T\|_F^2.$

To distinguish structured signal and noise, we need a clear measure of the relative strength of $\T$ and $\Q$. The most natural idea would be to use the nuclear norm. However, it is difficult to achieve a target VAF level with nuclear norms as shown in Equation~\ref{eq:vaf_x}. 
Instead, we use the Frobenius norm (sum of squared singular values) to proxy the relative component strength, as it enters the VAF calculation directly.
We draw $\T$ and $\Q$ simultaneously from the SVD of a single random $N \times P$ matrix with i.i.d. $\mathcal{N}(0,1)$ entries. The first $R$ left singular vectors, each multiplied by its corresponding singular value, form the columns of $\T$; the next $S$ left singular vectors, each multiplied by its corresponding singular value, form the columns of $\Q$. $\T$ is then rescaled so that $\|\T\|_F^2$ matches the Frobenius norm as used in Section~\ref{sec:sim1}. $\Q$ is rescaled to satisfy the target VAF. This simultaneous draw guarantees the orthogonality between signal and non-predictive components $\Qt\T = \mathbf{0}$ and $\Tt\T, \Qt\Q$ to be diagonal.

The matrices $\W\dims{P}{R}$ and $\K\dims{P}{S}$ are obtained as the first $R$ and next $S$ columns of the orthonormal factor from a QR decomposition of a single random $P \times (R+S)$ matrix. This ensures they are orthogonal to each other ($\Kt \W = \mathbf{0}$) and satisfy:
$\Wt\W = \IR, \quad \Kt\K = \IS.$
These properties simplify the Frobenius norm calculations in \eqref{eq:vaf_x}, as $\|\T\Wt\|_F^2 = \|\T\|_F^2$ and $\|\Q\Kt\|_F^2 = \|\Q\|_F^2$.

Regarding the test data, the random noise $\Extest$, $\Eytest$ is generated same as in Section~\ref{sec:sim1}. The latent scores $\T_\text{test}\dims{N_\text{test}}{R}$ and $\Q_\text{test}\dims{N_\text{test}}{S}$ were drawn independently, while the same loading matrices $\W$, $\Py$, and $\K$ from the training data is reused. This ensures that the test data share the same structural parameters as the training data, while having independent latent components. Results will be aggregated across the 100 Monte Carlo replications for each parameter combination.


\subsection{Results}

In Simulation~2, we examined the performance of PcovRnnp (with $\lse$ and $\Y$-loss fixed) against Ridge regression in a setting where the error term contains both a low-rank structured component and random Gaussian noise, in addition to the low-rank signal shared between $\X$ and $\Y$. We used the same performance evaluation metrics as in Section~\ref{sec:sim1}.

All figures in this section follow a consistent layout: The columns are grouped by the number of responses $(Q = 1, 5, 10)$, forming three blocks. The first block refers to univariate responses whereas the other two block refer to multivariate responses. Inside each block, the left panel displays results aggregated from conditions where the number of observations is more than the number of predictors ($N > P$). In contrast, the right panel displays results aggregated from conditions where the 
number of observations is less than the number of predictors ($N < P$). The rows refer to VAF levels. 

\newpage

\begin{figure}[htbp]
  \centering
  \includegraphics[width=1\linewidth]{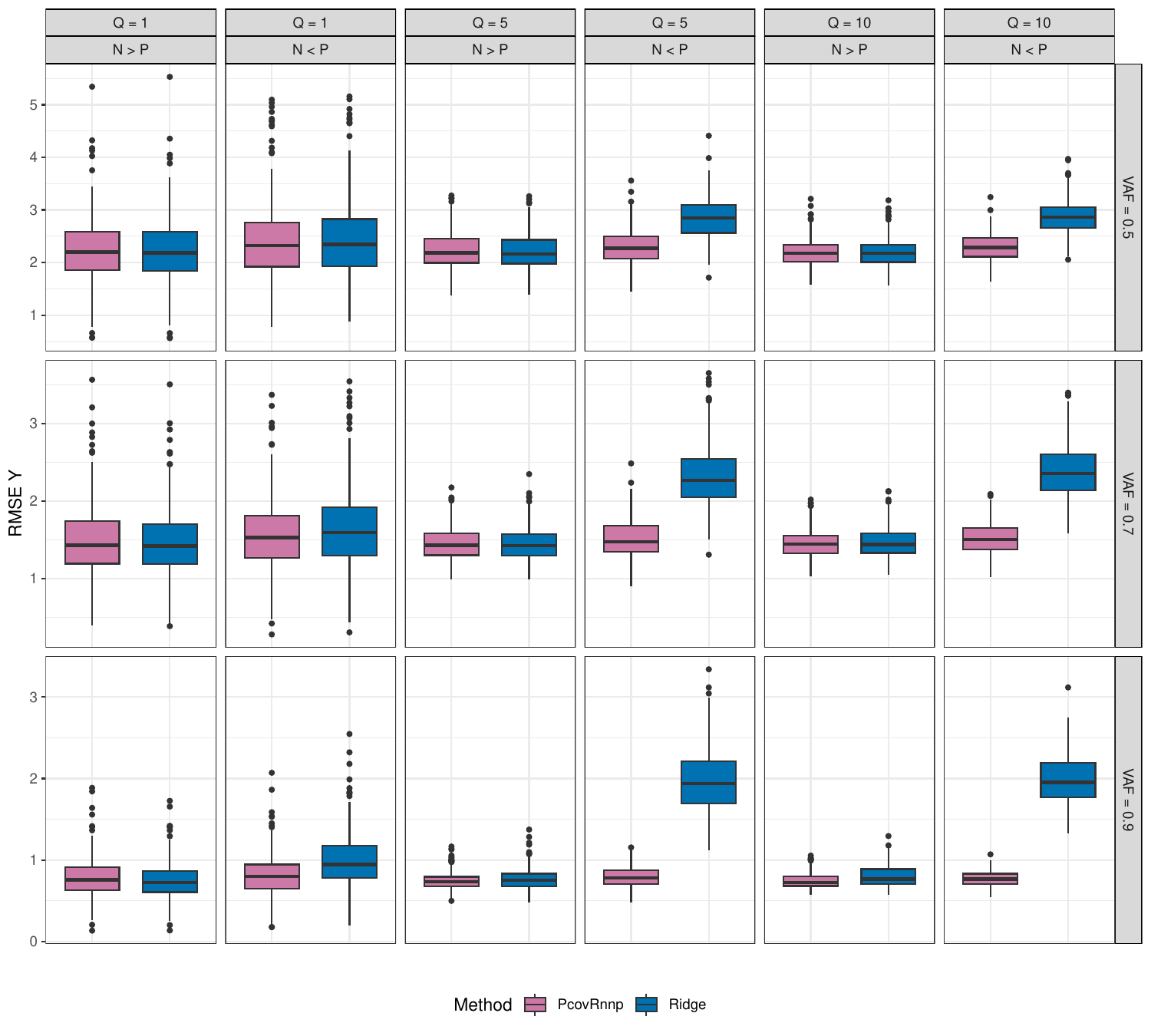}
  \caption{\textit{Root Mean Squared Error of $Y$ in Simulation~2.}}
  \label{fig:sim2_rmse_y}
  \vspace{0.2cm}
  \begin{minipage}{\linewidth}
    \small
    \textit{Note.} PcovRnnp: PcovRnnp using the 1-standard-error rule ($\lse$) targeting only $\Y$ 
    as CV objective.
    Ridge: Ridge regression using $\lse$ rules. 
    Test sample size $N_{\text{test}} = 1000$.
    Results are averaged over $100$ repetitions.
  \end{minipage}
\end{figure}

Figure~\ref{fig:sim2_rmse_y} shows the results regarding model performance in predicting outcomes. Regarding prediction error, nearly all PcovRnnp variants achieve lower RMSE on $\Y$ than Ridge regression in high-dimensional settings ($N < P$) with multivariate responses. This predictive advantage becomes stronger as VAF increases. In low-dimensional settings ($N > P$) or with a univariate response, the two methods perform comparably.

\newpage

\begin{figure}[htbp]
  \centering
  \includegraphics[width=1\linewidth]{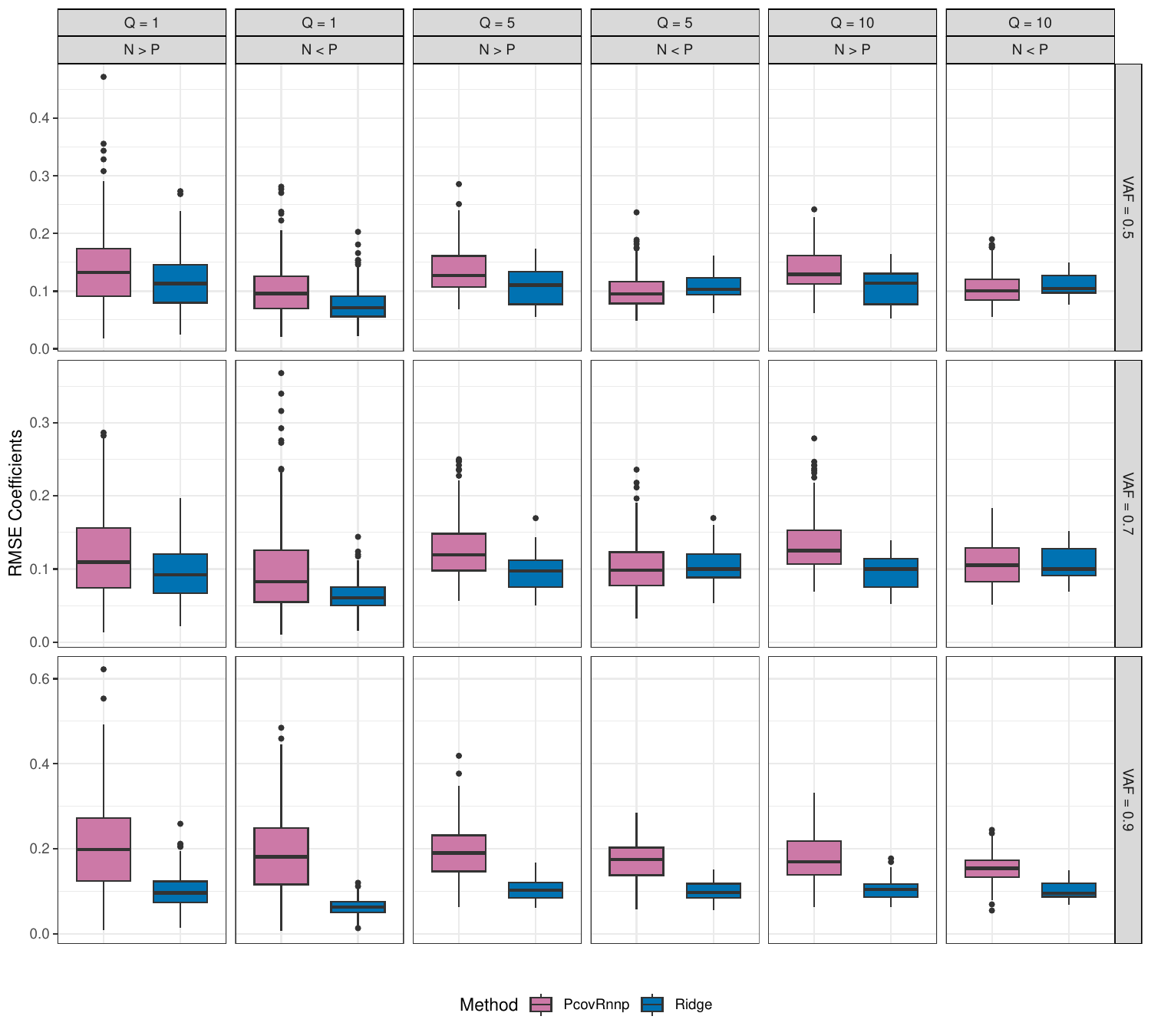}
  \caption{\textit{Root Mean Squared Error of Coefficients in Simulation~2.}}
  \label{fig:sim2_rmse_coef}
  \vspace{0.2cm}
  \begin{minipage}{\linewidth}
    \small
    \textit{Note.} PcovRnnp: PcovRnnp using the 1-standard-error rule ($\lse$) targeting only $\Y$ 
    as CV objective.
    Ridge: Ridge regression using $\lse$ rules.     
    Test sample size $N_{\text{test}} = 1000$.
    Results are averaged over $100$ repetitions.
  \end{minipage}
\end{figure}

Figure~\ref{fig:sim2_rmse_coef} shows the results regarding model performance in recovering coefficients. PcovRnnp tends to yield higher or similar RMSE relative to Ridge across most scenarios, with the gap narrowing in multivariate settings with low VAF.

\newpage

\begin{figure}[htbp]
  \centering
  \includegraphics[width=1\linewidth]{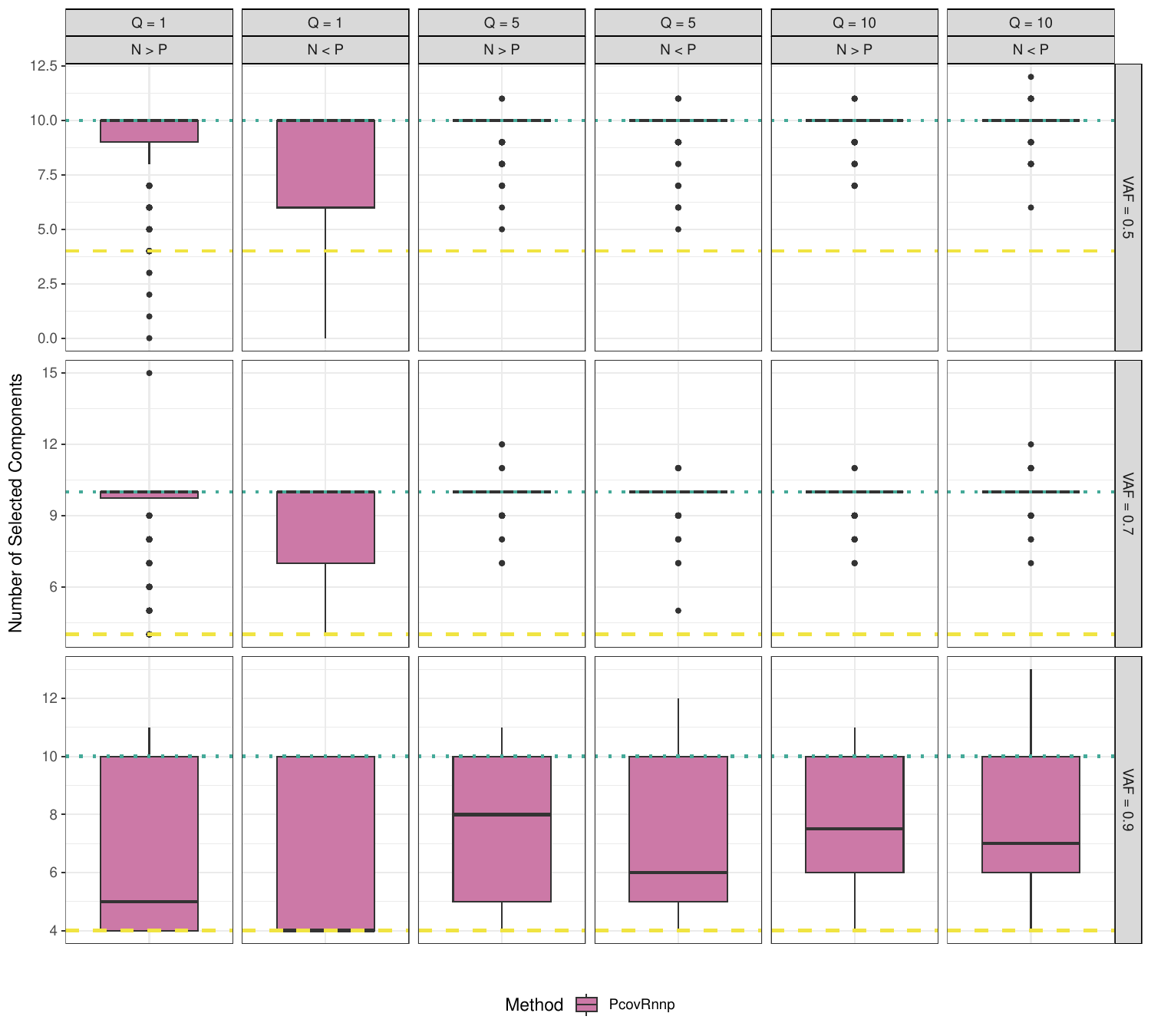}
  \caption{\textit{Number of Components Selected by PcovRnnp in Simulation~2.} }
  \label{fig:sim2_ncomp}
  \vspace{0.2cm}
  \begin{minipage}{\linewidth}
    \small
    \textit{Note.} PcovRnnp: PcovRnnp using the 1-standard-error rule ($\lse$) targeting only $\Y$ 
    as CV objective.
    Ridge: Ridge regression using $\lse$ rules. 
    Test sample size $N_{\text{test}} = 1000$.
    The horizontal dashed line refers to True rank $R = 4$.
    The horizontal dotted line refers to total predictor rank $R + S = 10$, where the true predictive rank $R = 4$ and non-predictive structure rank $S = 6$.
    Results are averaged over $100$ repetitions.
  \end{minipage}
\end{figure}

Figure~\ref{fig:sim2_ncomp} shows the results regarding component selection among PcovRnnp variants. PcovRnnp generally selects the number of components between the true signal rank $R = 4$ and the total rank $R + S = 10$. In univariate settings with high VAF, selections cluster near $R = 4$; in multivariate settings or under low VAF, they shift toward $R + S = 10$. This finding suggests that PcovRnnp absorbs the non-predictive components when the signal is weak or the response is complex.

\newpage
\section{Application: CERQ Dataset}
\label{sec:app1}


The dataset is obtained from a study on the psychometric properties of the Cognitive Emotion Regulation Questionnaire \citep[CERQ;][]{garnefski_cognitive_2007} in the general adult population . 
The predictors are 36 CERQ items consisting of 9 different subscales, each representing a cognitive or emotional coping strategy. The outcome variable is the sum score of the 16-item depression subscale  derived from the Symptoms Check List \citep[SCL;][]{arrindell_scl-90_1986} measured one year later. 
Due to a skewed distribution on the outcome, the outcome has been log-transformed prior to the analysis, resulting in a range of 2.77 to 4.38.
Due to dropout, we used listwise deletion same as described in \citet{pratiwi_predictive_2023}. Consequently, the sample size was reduced from 297 to 240.

    


\subsection{Comparative Evaluation Framework}
To evaluate the predictive performance of PcovRnnp against established methods, we apply 10-fold cross-validation repeated 10 times (100 iterations in total) to this dataset. Four methods are compared: PcovRnnp, Ridge regression, Sparse Principal Covariate Regression \citep[SPCovR;][]{van_deun_obtaining_2018}, and Sparse Partial Least Squares 
\citep[SPLS;][]{chun_sparse_2010}.
For PcovRnnp, $\alpha$ is estimated by maximum likelihood and $\lambda$ is selected via 10-fold  nested cross-validation, with 1-standard-error rule used for prediction. For SPCovR, we adopt the $\alpha$ by PcovRnnp since both methods use the same maximum-likelihood approach. We used a grid search to select the number of components (between 1 and 9) and $\lambda$. $\lse$ is used for prediction. For Ridge regression, we employ \texttt{cv.glmnet} \citep{friedman_glmnet_2008} with 10-fold cross-validation to select $\lambda$ and use $\lse$ for prediction. For SPLS, we evaluate component numbers between 1 and 9 and sparsity levels from 0.1 to 0.9, selecting optimal values via 10-fold cross-validation.
Performance is evaluated using out-of-sample RMSE across all response variables.

\subsubsection{Results}
Across 100 cross-validation iterations, the four methods demonstrated comparable predictive accuracy (see Table~\ref{tab:CERQ_results}). PcovRnnp, Ridge, and SPCovR each yielded an RMSE of $0.31 \pm 0.05$, with SPLS producing a marginally lower RMSE of $0.30 \pm 0.05$. 

Given this similar performance, we see that PcovRnnp offers a beneficial balance of interpretability, parsimony, and alignment with the underlying theoretical structure. While Ridge regression performs competitively, it does not explicitly estimate a latent structure; interpretation relies on the full set of individual coefficients, which can be challenging to navigate in theoretically structured predictor spaces. PcovRnnp provides an alternative by summarizing the space through a small number of interpretable components ($4.17 \pm 0.99$ on average). 
SPCovR shares this component-based approach but requires on average two more components ($6.51 \pm 1.87$) to achieve a similar prediction, suggesting a comparatively less parsimonious representation. 

Overall, PcovRnnp maintains competitive predictive accuracy while offering a more parsimonious solution than SPCovR, a component structure more reflective of the instrument's design than SPLS, and a dimension-reduced alternative to Ridge. This makes it a suitable option for psychological applications where interpreting data's latent structure is as important as predicting outcomes.

\begin{table}[htbp]
  \centering
  \begin{threeparttable} 
    \caption{\textit{Performance Comparison for CERQ Dataset}}
    \label{tab:CERQ_results}
    \small
    \begin{tabular}{lccc}
    \toprule
    Method & RMSE ($M \pm sd$) & Ncomp ($M \pm sd$) & $\alpha$ ($M \pm sd$) \\
    \midrule
    PcovRnnp & $0.31 \pm 0.05$ & $4.17 \pm 0.99$ & $0.97 \pm 0.00$ \\
    Ridge    & $0.31 \pm 0.05$ & -               & -               \\
    SPLS     & $0.30 \pm 0.05$ & $1.52 \pm 0.77$ & -               \\
    SPCovR   & $0.31 \pm 0.05$ & $6.51 \pm 1.87$ & $0.97 \pm 0.00$ \\
    \bottomrule
    \end{tabular}

    \begin{tablenotes}
      \footnotesize
      \item \textit{Note.} Ncomp: Number of latent components selected by the model.
      \item $\alpha$: weight parameter of PcovRnnp and SPCovR.
    \end{tablenotes}
  \end{threeparttable}
\end{table}

\subsection{Interpret PcovRnnp on CERQ}

To illustrate the practical usage of PcovRnnp, we fitted one PcovRnnp model on the CERQ 
dataset using 10-fold CV\@. This model selected $\lse = 0.214$, yielding an in-sample $R^2 = 0.18$ and reducing 36 CERQ items 
into four components that summarize the CERQ score information while simultaneously predicting 
depression. Table~\ref{tab:CERQ_combined} displays the component weights, response 
loadings, and coefficients.

Component 1 (loading $= 1.143$) reflects deliberate cognitive reappraisal. It focuses high weights on refocus planning and positive reappraisal. Both strategies
involve effortful, goal-directed restructuring of how one thinks about a stressor. The
modest response loading suggests that this reappraisal tendency is also deployed reactively
under elevated distress.

Component 2 (loading $= 3.113$) is the strongest predictor of depression. It is characterized by
high weights on catastrophizing and blaming others. Both strategies share a tendency to
exaggerate the perceived severity of the stressor while directing responsibility outward. The
large response loading supports that this dimension is the primary predictor of depression
severity in the model.

Component 3 (loading $= 0.406$) reflects attentional engagement with the stressor. It
contrasts rumination against strategies that redirect attention elsewhere, such as positive
refocusing, refocus planning, and putting the event into perspective. High scorers show
persistent, repetitive focus on the stressor rather than shifting toward more constructive
content. However, the response loading of this component is relatively small compared to components 1 and 2. 

Component 4 (loading $= 0.240$) captures the attribution direction of cognitive emotion
regulation. It assigns positive weights to self-blame, acceptance,
positive refocusing, and refocus planning - strategies that internalize thoughts. It assigns a negative weight to blaming others, one externalizing strategy. The response loading of the component is the lowest.

Regarding the regression coefficients linking individual items to depression, items from the catastrophizing and blaming others domains show the highest positive coefficients among all predictors — specifically, catastrophizing items 2, 3, and 4, blaming others items 1 through 4, self-blame item 1, and rumination items 1 and 2. This pattern indicates that individuals who more frequently employ catastrophizing or blaming others as coping strategies tend to report higher depression scores.

\begin{table}[htbp]
  \centering
  \caption{\textit{Component Weights, Response Loadings, and Regression Coefficients on CERQ}}
  \label{tab:CERQ_combined}
  \small
  \setlength{\tabcolsep}{8pt}
  \begin{threeparttable}
    \begin{tabular}{lrrrrr}
    \toprule
    {Variable\tnote{a}} & {Comp.\ 1} & {Comp.\ 2} & {Comp.\ 3} & {Comp.\ 4} & {Coefficient\tnote{c}} \\
    \midrule
    Self-blame 1              &  2 &  6 & -2 & \textbf{16} & \textbf{25.3} \\
    Self-blame 2              &  3 & -1 & -6 & \textbf{17} &  0.6 \\
    Self-blame 3              &  4 &  0 & -2 & \textbf{11} &  6.1 \\
    Self-blame 4              &  3 &  3 &  0 & \textbf{19} & 17.7 \\
    \addlinespace
    Acceptance 1              &  4 &  1 &  3 &  6 & 11.1 \\
    Acceptance 2              &  4 &  0 &  2 &  6 &  6.5 \\
    Acceptance 3              &  2 &  4 &  5 &  9 & 17.7 \\
    Acceptance 4              &  4 &  3 &  2 &  4 & 14.9 \\
    \addlinespace
    Rumination 1              &  3 &  7 & \textbf{-5} & -2 & \textbf{24.1} \\
    Rumination 2              &  3 &  6 & \textbf{-6} & -3 & \textbf{20.0} \\
    Rumination 3              &  4 &  4 & \textbf{-9} & -2 & 12.2 \\
    Rumination 4              &  4 &  4 & \textbf{-7} & -8 & 12.6 \\
    \addlinespace
    Positive Refocusing 1     &  3 & -4 & \textbf{10} & \textbf{-6} & -6.3 \\
    Positive Refocusing 2     &  3 & -2 & \textbf{12} & \textbf{-7} &  0.4 \\
    Positive Refocusing 3     &  3 & -3 & \textbf{12}& \textbf{-5} & -1.6 \\
    Positive Refocusing 4     &  4 & -4 &  \textbf{9} & \textbf{-6} & -7.8 \\
    \addlinespace
    Refocus Planning 1        &  \textbf{5} & -2 & \textbf{-5} & -6 & -3.4 \\
    Refocus Planning 2        &  \textbf{5} & -1 & \textbf{-7} & -7 & -1.9 \\
    Refocus Planning 3        &  \textbf{5} &  0 & \textbf{-9} & -4 & -0.3 \\
    Refocus Planning 4        &  \textbf{5} & -1 & \textbf{-7} & -7 & -2.5 \\
    \addlinespace
    Positive Reappraisal 1    &  \textbf{5} & -4 & -5 & -2 & -10.5 \\
    Positive Reappraisal 2    &  4 & -2 & -3 & -7 &  -5.0 \\
    Positive Reappraisal 3    &  \textbf{5} & -6 & -4 &  2 & -14.1 \\
    Positive Reappraisal 4    &  \textbf{5} & -5 & -3 & -3 & -11.0 \\
    \addlinespace
    Putting into Perspective 1 &  4 & -2 &  \textbf{8} &  3 &  3.0 \\
    Putting into Perspective 2 &  3 & -3 &  \textbf{8} &  7 & -0.8 \\
    Putting into Perspective 3 &  4 & -6 &  \textbf{5} &  8 & -9.1 \\
    Putting into Perspective 4 &  4 & -4 &  \textbf{7} &  4 & -3.9 \\
    \addlinespace
    Catastrophizing 1         &  1 &  4 &  2 &  5 & 14.9 \\
    Catastrophizing 2         &  1 & \textbf{11} &  3 &  3 & \textbf{36.4} \\
    Catastrophizing 3         &  1 &  \textbf{8} &  5 &  3 & \textbf{29.9} \\
    Catastrophizing 4         &  2 & \textbf{10} &  2 & -2 & \textbf{33.4} \\
    \addlinespace
    Blaming Others 1          &  1 &  \textbf{6} &  6 & -4 & \textbf{22.5} \\
    Blaming Others 2          &  1 &  \textbf{9} &  6 & -6 & \textbf{30.2} \\
    Blaming Others 3          &  3 &  \textbf{7} &  5 & -7 & \textbf{24.3} \\
    Blaming Others 4          &  1 &  \textbf{7} &  7 & -8 & \textbf{25.4} \\
    \midrule
    Response                  &    &    &    &    &      \\
    \addlinespace
    {Depression\tnote{b}}     & 1.143 & 3.113 & 0.406 & 0.240 & \\
    \bottomrule
    \end{tabular}
    \begin{tablenotes}[flushleft]
      \small
      \item \textit{Note.} \item[a]All component weights and regression coefficients are
        multiplied by $10^{-3}$; divide by 1{,}000 to recover the original scale.
      \item[b] The Depression row reports the response loadings (i.e., the regression
        coefficients from components to depression). 
       \item[c]  The coefficient of each predictor
        can be expressed as the inner product of its component weights and the response
        loadings. For example, the coefficient for Self-blame~1 is obtained as
        $0.002 \times 1.143 + 0.006 \times 3.113 + ({-0.002}) \times 0.406
         + 0.016 \times 0.240 \approx 0.024$
        (${\approx}25.3 \times 10^{-3}$ after rounding the weight entries).
    \end{tablenotes}
  \end{threeparttable}
\end{table}

\section{Application: HCP Dataset}
\label{sec:app2}
\subsection{Data description}
Neuroimaging data are increasingly used to understand variation in human behaviours in applied psychological research. For the second application we acquired data from the Human Connectome Project (HCP) Young Adult dataset \citep{van_essen_wu-minn_2013}. The Human Connectome Project set out to map the 'human connectome', a database of functional connections between brain regions under different conditions (e.g., resting-state, functional, or structural connectivity). The HCP provides high-dimensional datasets that are often used in Brain-Wide Association studies (BWAS) \citep[]{marek_reproducible_2022, smith_positive-negative_2015}. BWAS are studies that examine the associations between human brain functioning and different phenotypes (e.g. executive function, episodic memory). The aim of this application study is to assess model performance in an applied neuroimaging context, using brain-derived measures to obtain predictions across a range of phenotypes. 
Recent large-scale neuroimaging analysis studies investigated the replicability of effect sizes found in neuroscience and reached opposite conclusions regarding replicability and sample size requirements \citep[]{spisak_multivariate_2023, marek_reproducible_2022}. While \citet{marek_reproducible_2022} demonstrated that BWAS effect sizes are in-sample inflated in smaller samples, \citet{spisak_multivariate_2023} showed that in-sample inflation can be reduced by appropriate cross-validation procedures. Given this ongoing debate, we assess both in- and out-of-sample performance to evaluate the replicability of findings.
A total of 985 participants, ranging between ages 22 and 35 (\textit{M}=28.8, \textit{SD}=3.59), with completed data were used in the further analysis of this study. HCP participants underwent several behavioural tests that are part of the National Institutes of Health (NIH) toolbox battery and several non-toolbox behavioural measures, such as delay discounting and emotion recognition \citep[]{barch_function_2013}.\footnote{The NIH Toolbox consists of a comprehensive set of neuro-behavioural measurements on the domains of cognition, emotion, motor function and sensation. See \url{http://www.nihtoolbox.org} for more information} The following 7 behavioural measures are used as outcome variables: age; cognitive ability; episodic memory; fluid intelligence; cognitive flexibility; inhibition and anger affect. All measures are age-adjusted scale scores, except for anger affect. The non-cognitive affective questionnaire denoted here as anger affect is a questionnaire measure that is not age adjusted while all other measures are behavioural tasks. Cognitive ability is a composite measure of cognitive function that includes all other NIH toolbox cognition measures. More information on the variables can be found in \citet[2225]{van_essen_human_2012}. For this application we used resting-state functional MRI (rs-fmri) data as brain features. More specifically, we used the first 100 components of a spatial independent component analysis (ICA) performed on each individual's resting-state time-series \citep[]{beckmann_group_2009}. Each component represents a spatial map of functionally connected brain regions. For each participant, a partial correlation was calculated between the timeseries of each of the pairs of the 100 brain regions\citep[]{van_essen_human_2012, glasser_human_2016}, resulting in a \textit{N} $\times$ 4950 matrix. This matrix was used to assess the association between brain connectivity and the various phenotypes.

\subsection{Performance} 

\subsubsection{Comparative Evaluation Framework}
To assess the performance of PcovRnnp against other methods, models are evaluated in terms of predictive performance and (non-)replicability. A simple replication framework is implemented where data is split into a calibration (\textit{N}=600) and a validation dataset (\textit{N}=385). 
First, model performance is examined on the calibration dataset. Within the calibration dataset 10-fold cross-validation is performed and a cross-validated $R^2$ is derived for every fold. There are various ways for calculating $R^2$. The formula we use to calculate $R^2$ is defined as:  \[
R^2 = 1-\frac{\text{RSS}}{\text{TSS}},
\] where RSS is the residual sums of squares and TSS the total sum of squares. \\ 
Secondly, to evaluate replicability, models are assessed on how well they perform on unseen data from the same population. Hence, models are trained on the full calibration set and a `nested 10-fold cross-validation` is performed to obtain the optimal penalty parameter value $\lambda_{1se}$, using the one standard error rule.\footnote{Nested refers here to the fact that we do not perform 10-fold cross-validation on the complete dataset. 10-fold cross-validation is only performed to obtain tuned hyperparameter estimates} The obtained parameter estimates are used to acquire predictions on the validation dataset. These validation results are then contrasted against the cross-validated calibration results.
Finally, this framework is applied across all methods twice: For every method, models are obtained by fitting each response individually and by fitting all response variables jointly. Corresponding results (e.g. $R^2$ values) are labelled either individual or joint accordingly.

\subsubsection{Hyperparameter tuning}
Similar to the previous application study (Section \ref{sec:app1}), we intended to compare the following methods: PcovRnnp, ridge regression, Sparse Principal Covariate Regression (SPCovR) and Sparse Partial Least Squares (SPLS). All computations were performed on the ALICE (Academic Leiden Interdisciplinary Cluster Environment) High Performance Computing facility provided by Leiden University, using 1 CPU node with EPYC 9554P (Zen4) and dual Xeon Gold 6126 (Skylake) processors and 384 GB RAM per node. A walltime of 168 hours was allocated per job. SPCovR runs into a time-out issue as it fails to converge within the allocated walltime (168 hours) on the computer cluster. In contrast, PcovRnnp completed the 10-fold cross-validation within $\approx$ 1 hour, ridge in $\approx$ 5 minutes and SPLS within $\approx$ 40 hours.\footnote{All methods adhered to the same resources and procedure of the computer cluster. All jobs were submitted simultaneously and ran in two waves (jobs 1-5 and jobs 6-10). For the entire cross-validation, folds ran in parallel across cluster nodes. Each cross-validation fold was allocated to a node, with the following resources: billing=4, cpu=1, memory=15G.} Due to this high computational complexity exhibited by SPCovR, the method was ommited from the further analysis of results. 
Given the specific method, hyperparameters are tuned. For PcovRnnp, $\alpha$ is estimated by maximum likelihood and $\lambda $ is selected via 10-fold nested cross-validation. The one standard error rule is used to select the $\lambda_{1se}$ for prediction. Similarly, for ridge regression $\lambda$ is selected with 10-fold nested cross-validation and based on the one standard error rule, $\lambda_{1se}$ is used for prediction. Finally, for SPLS, component numbers (\textit{K}) between 1 and 100 and sparsity levels ($\eta$) from 0.1 and 0.9 are evaluated and optimal values selected by 10-fold nested cross-validation.

\subsection{Results}

\subsubsection{Comparison of performance of methods}
First, we compare the results for the three methods in terms of predictive performance for the individual and for joint response models. Secondly, we interpret the results in terms of replicability. The results for the individual response models for the calibration and validation dataset are shown respectively in \autoref{tab:r2_uni_combined}\textit{A} \& \textit{B}. For the joint response models the results for the calibration and validation dataset set are shown respectively in \autoref{tab:r2_multi_combined}\textit{A} \& \textit{B}. We denote $R^2$ as a measure of predictive performance, in terms of predictive accuracy. $R^2$ represents the out-of-sample proportion of variance in the outcome explained by the model, where higher values indicate better predictive performance. A negative out-of-sample $R^2$ value indicates that the model performs worse than a model that only predicts the mean. Variables were considered predictable if the cross-validated $R^2$ value exceeded zero beyond one standard deviation across folds (i.e. zero did not fall within the \textit{M} $\pm$ \textit{SD} range). Results were considered replicable if the validation $R^2$ values fell within the \textit{M} $pm$ \textit{SD} range of the calibration set.

\begin{table}[htbp]
  \centering
  \begin{threeparttable}
    \caption{\textit{Individual $R^2$ for HCP Calibration and Validation Datasets: PcovRnnp, Ridge, and SPLS}}
    \label{tab:r2_uni_combined}
    \small
    \begin{tabular}{lrrr}
      \toprule
      \textbf{Variable}
        & \textbf{PcovRnnp}
        & \textbf{Ridge}
        & \textbf{SPLS} \\
      \midrule
      \multicolumn{4}{l}{\textit{A: Calibration Dataset}} \\
      \midrule
      Age
        & $0.144 \pm 0.050$ & $0.248 \pm 0.065$ & $0.276 \pm 0.090$ \\
      Cognitive ability
        & $0.105 \pm 0.060$ & $0.184 \pm 0.073$ & $0.209 \pm 0.088$ \\
      Fluid Intelligence
        & $0.031 \pm 0.060$ & $0.045 \pm 0.062$ & $0.029 \pm 0.176$ \\
      Inhibition
        & $-0.009 \pm 0.014$ & $0.003 \pm 0.016$ & $-0.136 \pm 0.102$ \\
      Cognitive flexibility
        & $-0.026 \pm 0.052$ & $-0.008 \pm 0.055$ & $-0.030 \pm 0.068$ \\
      Episodic memory
        & $0.010 \pm 0.035$ & $0.024 \pm 0.023$ & $-0.025 \pm 0.114$ \\
      Anger affect
        & $-0.027 \pm 0.020$ & $-0.022 \pm 0.020$ & $-0.055 \pm 0.086$ \\
      \midrule
      \multicolumn{4}{l}{\textit{B: Validation Dataset}} \\
      \midrule
      Age
        & $0.151$ & $0.221$ & $0.177$ \\
      Cognitive ability
        & $0.146$ & $0.225$ & $0.189$ \\
      Fluid Intelligence
        & $0.093$ & $0.104$ & $0.059$ \\
      Inhibition
        & $-0.005$ & $0.008$ & $-0.100$ \\
      Cognitive flexibility
        & $-0.015$ & $0.002$ & $-0.020$ \\
      Episodic memory
        & $0.025$ & $0.047$ & $0.026$ \\
      Anger affect
        & $-0.012$ & $-0.007$ & $-0.138$ \\
      \bottomrule
    \end{tabular}
    \begin{tablenotes}
      \footnotesize
      \item \textit{Note.} \textit{A. Calibration Dataset}: values are $R^2$ (mean across cross-validation folds) $\pm$ $SD$ (standard deviation across folds) on the calibration dataset. \textit{B. Validation Dataset}: $R^2$ values on the held-out validation set. Both model the responses individually.
      \item PcovRnnp: Principal Covariates Regression with Nuclear Norm Penalty; Ridge: Ridge Regression; SPLS: Sparse Partial Least Squares.
    \end{tablenotes}
  \end{threeparttable}
\end{table}

Comparing the cross-validated individual $R^2$ values shown in \autoref{tab:r2_uni_combined}\textit{A}, only two responses were predictive across all methods: age and cognitive ability. All other variables were not reliably predictive by any method as zero fell within the $M \pm SD$ range, suggesting predictions had no consistent positive predictive accuracy across folds. Age might be considered a reference variable, resulting in a relatively high $R^2$, while the higher $R^2$ value for cognitive ability is expected given that it is a composite of its subcomponents. Considering these two response variables, the $R^2$ values for PcovRnnp is lower (age: 0.144 $\pm$ 0.050, cognitive ability: 0.105 $\pm$ 0.060), than both ridge regression and SPLS. However, given the overlap between the upper end of PcovRnnp's \textit{M} $\pm$ \textit{SD} range and the lower end of the other two methods, no concrete conclusions can be drawn that these methods outperform PcovRnnp when responses are modelled individually. \\ 

\begin{table}[htbp]
  \centering
  \begin{threeparttable}
    \caption{\textit{Joint $R^2$ for HCP Calibration and Validation Datasets: PcovRnnp, Ridge, and SPLS}}
    \label{tab:r2_multi_combined}
    \small
    \begin{tabular}{lrrr}
      \toprule
      \textbf{Variable}
        & \textbf{PcovRnnp}
        & \textbf{Ridge}
        & \textbf{SPLS} \\
      \midrule
      \multicolumn{4}{l}{\textit{A: Calibration Dataset}} \\
      \midrule
      Age
        & $0.131 \pm 0.051$ & $0.078 \pm 0.034$ & $0.040 \pm 0.084$ \\
      Cognitive ability
        & $0.111 \pm 0.069$ & $0.058 \pm 0.046$ & $0.200 \pm 0.103$ \\
      Fluid Intelligence
        & $0.057 \pm 0.068$ & $0.017 \pm 0.046$ & $0.116 \pm 0.123$ \\
      Inhibition
        & $0.007 \pm 0.028$ & $0.002 \pm 0.017$ & $-0.029 \pm 0.084$ \\
      Cognitive flexibility
        & $0.003 \pm 0.069$ & $-0.003 \pm 0.058$ & $-0.004 \pm 0.137$ \\
      Episodic memory
        & $0.039 \pm 0.031$ & $0.021 \pm 0.018$ & $-0.049 \pm 0.067$ \\
      Anger affect
        & $-0.027 \pm 0.021$ & $-0.021 \pm 0.018$ & $-0.042 \pm 0.033$ \\
      \midrule
      \multicolumn{4}{l}{\textit{B: Validation Dataset}} \\
      \midrule
      Age
        & $0.131$ & $0.080$ & $0.067$ \\
      Cognitive ability
        & $0.140$ & $0.077$ & $0.216$ \\
      Fluid Intelligence
        & $0.096$ & $0.054$ & $0.148$ \\
      Inhibition
        & $0.021$ & $0.009$ & $-0.083$ \\
      Cognitive flexibility
        & $0.020$ & $0.008$ & $-0.066$ \\
      Episodic memory
        & $0.050$ & $0.034$ & $0.008$ \\
      Anger affect
        & $-0.002$ & $-0.007$ & $-0.014$ \\
      \bottomrule
    \end{tabular}
    \begin{tablenotes}
      \footnotesize
      \item \textit{Note.}\textit{A. Calibration Dataset}: values are $R^2$ (mean across cross-validation folds) $\pm$ $SD$ (standard deviation across folds) on the calibration dataset. \textit{B. Validation Dataset}: $R^2$ values on the held-out validation set. Both model the responses jointly.
      \item PcovRnnp: Principal Covariates Regression with Nuclear Norm Penalty; Ridge: Ridge Regression; SPLS: Sparse Partial Least Squares.
      \item PCovRnnp; \textit{A}: $\alpha = 0.997\ (SD = 5.18 \times 10^{-5})$, ncomp $= 111\ (SD = 35.1)$;  \textit{B}: $\alpha = 0.997$, ncomp $= 132$.
      \item SPLS; \textit{A}: $\eta = 0.18\ (SD = 0.09)$, ncomp $= 4\ (SD = 1.05)$;  \textit{B}: $\eta = 0.1$, ncomp $= 5$.
    \end{tablenotes}
  \end{threeparttable}
\end{table}
When responses are modelled jointly, shown in \autoref{tab:r2_multi_combined}\textit{B}, the cross-validated joint $R^2$ values for PcovRnnp decreased slightly for age (0.131 $\pm$ 0.051) while remaining similar for cognitive ability (0.111 $\pm$ 0.069). In contrast, joint modelling of responses led to a substantial decline in $R^2$ for both ridge regression and SPLS, with age becoming non-predictable for SPLS (0.040 $\pm$ 0.084). A possible reason for this decline might lie in differences in hyperparameter tuning. For individual models, a penalty parameter is optimised for each response variable seperately. For joint models, a single penalty parameter is shared across all outcomes, resulting in a compromise between different levels of regularization and thereby reducing the overall predictive accuracy. For PcovRnnp, the tuned $\alpha$ parameter converged to approximately 1 (0.997), indicating that the model essentially reduced to Principal Component Analysis (PCA) and the response variables had little influence. This might explain why PcovRnnp shows more stability in $R^2$ compared to the other methods. Overall, results suggest that PcovRnnp performs comparable to the other methods in the individual case and shows less decline in predictive accuracy when responses are modelled jointly. \\
Next, we compared the results of the calibration and validation datasets. The data from the validation set are treated as an independent sample from the same population to assess replicability. Individual and joint $R^2$ values for the validation set are given in \ref{tab:r2_uni_combined}\textit{B} and \ref{tab:r2_multi_combined}\textit{B} respectively. Three results fell outside the \textit{M} $\pm$ 1 \textit{SD)} range of the calibration dataset: fluid intelligence for PcovRnnp in the individual model (0.093, calibration range: -0.029 to 0.091), anger affect for PcovRnnp in the joint model (-0.002, calibration range: -0.048 to -0.006) and age for SPLS in the individual model (0.177, calibration set: 0.186, 0.366]). For PcovRnnp these deviances are small and in the positive direction, indicating slightly better predictions in the validation set. For all methods, both individual and joint models were replicable for age and cognitive ability with the exception of the joint SPLS model. Overall, these results suggest that estimates are replicable for both the individual and joint models. \\


\section{Discussion}
\label{sec:discussion}

In this article, we have introduced PcovRnnp, an extension of the PCovR framework. By incorporating a nuclear norm penalty, PcovRnnp allows for simultaneously selecting the number of components and estimating regularized coefficients instead of extensive grid-search over the number of components required by other PCovR methods. We have provided a comprehensive theoretical analysis establishing that PcovRnnp is a mathematically consistent and correct extension of PCovR. Furthermore, we developed an efficient algorithm to find the optimal solution.

Our simulation studies supported the advantage of PcovRnnp over existing approaches. 
In Simulation~1, where the error term contained only random Gaussian noise and the predictors had a low-rank structure, PcovRnnp consistently outperformed Ridge regression in predicting outcomes in high-dimensional settings with multivariate responses — that is, settings in which several response variables are modeled jointly under a single penalty parameter. Regarding the penalty parameter, we found that using the 1-standard-error rule yielded sparser and more accurate coefficient structures compared to the minimum-error rule. Furthermore, combining the 1-standard-error rule with a cross-validation criterion focused on outcomes provided the most accurate recovery of the true rank. However, PcovRnnp tended to show higher or at least similar RMSE for coefficients compared to Ridge regression.
In Simulation~2, where the error term contained both random noise and non-predictive latent components, PcovRnnp maintained its advantage in predicting outcomes against Ridge regression while having higher or similar RMSE on coefficients compared to Ridge regression. Interestingly, the PcovRnnp model tended to select the number of components between the true signal rank ($R = 4$) and the total rank ($R + S = 10$). This finding suggests that PcovRnnp selects the structured noise components to reconstruct the predictors, although those noise components do not predict outcomes.

In the empirical application using the CERQ dataset, PcovRnnp demonstrated comparable predictive accuracy (RMSE on outcomes) to Ridge, SPLS, and SPCovR under 10-times repeated 10-fold cross-validation, while offering meaningful advantages in parsimonious component selection. On average, PcovRnnp selected two fewer components than SPCovR, yielding a more compact and interpretable model. Although SPLS extracted the fewest components ($1.52 \pm 0.77$), its components are optimized primarily for the predictive covariance between predictors and outcomes. Consequently, the SPLS latent components are difficult to interpret, limiting its utility for psychological research where theoretical explanation is as critical as predictive performance \citep{yarkoni_choosing_2017}.
PcovRnnp mitigates this limitation by jointly balancing variance explanation in predictors and outcome prediction in outcomes. This balance is reflected in the estimated weighting parameter $\hat{\alpha} \approx 0.97$ for both PcovRnnp and SPCovR, which indicates a strong emphasis on recovering the underlying predictor structure while retaining predictive accuracy.
Fitting a single PcovRnnp model to the CERQ dataset yielded a four-component solution that illustrates these properties. The model identified self-blame, rumination, and catastrophizing as predictive of depression, consistent with \citet{garnefski_cognitive_2007}, and additionally highlighted blaming others as a substantively meaningful predictor. These findings suggest that explicitly modeling the shared latent structure of the CERQ subscales is essential for extracting components that are both interpretable and predictive, which is an advantage that purely prediction-focused methods such as SPLS may not fully provide.

For the second empirical application, data from the Human Connectome Project (HCP) were employed to serve as an example of high-dimensional data often acquired in brain-wide association studies (BWAS). A comparison was made between modelling the responses separately (univariate) or jointly (multivariate). PcovRnnp showed to outperform the other methods in terms of predictive accuracy ($R^2$) for the multivariate models. Hence, for both Ridge and SPLS, jointly modeling all outcomes simultaneously comes at the cost of reduced predictive accuracy. This might suggest that PcovRnnp would be particularly beneficial in scenarios where predicting multivariate responses is desirable. This result is further evidenced by the fact that the number of components of the multivariate model is lower than that of some of the univariate models and suggests that the lower-dimensional solution is beneficial since the outcomes share a common predictor space. 
Furthermore, we note that the $\alpha$ hyperparameter for PcovRnnp for both models is close to 1 ($\approx$ 0.997), indicating that the models primarily minimize reconstruction error on the predictor matrix. Hence, only a small portion is reserved to minimize the prediction error for the outcome variable(s). This might suggest that for high-dimensional brain datasets, it is more important to accurately recover the structure of the connectivity matrix rather than to optimize the prediction of the outcome variables directly. 
Finally, we assessed replicability by evaluating model performance in a separate validation dataset. Recent studies have suggested that BWAS are non-replicable and would require thousands of participants to retrieve replicable results \citep[]{marek_reproducible_2022}. In contrast, our findings indicate that the validation results yield comparable or larger $R^2$ values than those obtained from the calibration set. This might suggest that the models do not overfit and that the results from the calibration set are not inflated. Similar to the findings of \citet{spisak_multivariate_2023}, our results show that, in general, the methods return replicable results.  


Despite the promising results, this article has several limitations that are worth consideration.
First, PcovRnnp does not show a clear advantage in recovering true coefficients compared to Ridge regression. One explanation is that the nuclear norm penalty promotes low-rank structure in the coefficient matrix but does not directly shrink individual coefficient entries toward zero like a Ridge penalty. Some research suggests that combining a nuclear norm penalty with an additional Ridge penalty could yield more accurate coefficient estimates \citep{chen_reduced_2013}. Incorporating such a composite penalty into PcovRnnp is a natural extension and could be implemented in future work.
Second, we would like to point out that, in Simulation~2, the ML approach to estimate the weight hyperparameter $\alpha$ might be theoretically inappropriate because the ML framework assumes error to be completely Gaussian \citep{vervloet_selection_2013}. Although the results from Simulation~2 demonstrate that PcovRnnp still performs reasonably well compared to Ridge regression, suggesting PcovRnnp's robustness for real-world data, a better choice of $\alpha$ may further improve model performance. For example, one could tune $\alpha$ via grid search over $[0.1, 0.2, ..., 0.9]$, though this becomes computationally expensive for large datasets. 
Third, the simulation and application examples focused mainly on continuous responses with a single block of predictors. However, in research, there is also interest in multiblock data, e.g., predicting depression symptoms from multiple questionnaires or predicting cognitive capacity based on both structural and functional brain imaging data.
Future research is suggested to incorporate additional regularizations to stabilize coefficient estimates, develop an alternative approach to estimate $\alpha$ that relaxes assumptions about noise, and extend PcovRnnp to model multiblock data.

To conclude, our theoretical and empirical results suggest that PcovRnnp offers an efficient and interpretable framework for simultaneously selecting latent components and predicting outcomes in psychological research. The components selected by PcovRnnp capture the structure of the predictors while naturally predicting the outcomes, making it well-suited for applications such as psychological questionnaires predicting symptom profiles. Furthermore, PcovRnnp is computationally efficient compared to existing PCovR methods — particularly for high-dimensional settings such as brain-imaging data, where grid-search over the number of components renders alternative approaches substantially slower.

We hope that the accompanying R package PcovRnnp (available in \hyperlink{https://github.com/I-N0-K/PcovRnnp}{https://github.com/I-N0-K/PcovRnnp}) makes this method accessible to applied researchers working with high-dimensional data across fields such as psychology, neuroscience, and beyond, and that it eases the interpretation of complex multivariate results in practice. 


\paragraph{Acknowledgements}
We appreciate Dr.Garnefski and Dr.Kraaij for access to the CERQ data.
This work was performed using the ALICE compute resources provided by Leiden University.

\paragraph{Funding Statement}
This publication is part of the project "Linking brain function and structure to phenotypes: does more data in fixed sample sizes lead to higher replicability?" of the research programme SGW Open Competition which is financed by the Dutch Research Council (NWO) under the grant 406.23.PPO.019.

\paragraph{Competing Interests}
The authors have no relevant financial or non-financial interests to disclose.

\paragraph{Data Availability Statement}
The R package PcovRnnp can be found at: \hyperlink{https://github.com/I-N0-K/PcovRnnp}{https://github.com/I-N0-K/PcovRnnp}.
The simulation and application scripts can be found at: \hyperlink{https://github.com/I-N0-K/pcovrnnpArticle}{https://github.com/I-N0-K/pcovrnnpArticle}. The HCP dataset can be found at: \hyperlink{https://www.humanconnectome.org/study/hcp-young-adult/data-releases}{https://www.humanconnectome.org/study/hcp-young-adult/data-releases}.

\paragraph{Author Contributions}
Conceptualization: K.L; M.dR. Software: K.L. Formal Analysis: K.L; L.V; W.W; M.dR. Investigation: K.L; L.V. Writing – original draft: K.L; L.V. Writing – review \& editing: K.L; L.V; W.W; M.dR. Supervision: W.W; M.dR. All authors approved the final submitted draft.

\printbibliography

\end{document}